\documentclass[longauth]{aa} 
\newcommand{\grvs}{$G_\mathrm{RVS}$}
\newcommand{\extgrvs}{$G_\mathrm{RVS}^{\rm{ext}}$}
\newcommand{\intgrvs}{$G_\mathrm{RVS}^{\rm{int}}$}
\newcommand{\refgrvs}{$G_\mathrm{RVS}^{\rm{ref}}$}

\newcommand{\RV}{$V_\mathrm{R}$}
\newcommand{\RVtr}{$V_\mathrm{R}^{\rm t}$} 
\newcommand{\RVref}{$V_\mathrm{R}^{\rm{ref}}$} 
\newcommand{\RVerr}{$\sigma_{V_\mathrm{R}}$}
\newcommand{\SRVCCD}{$sV_\mathrm{R}^{\mathrm{CCD}}$} 
\newcommand{\SRV}{$sV_\mathrm{R}^{\rm t}$} 

\newcommand{\FeH}{[Fe/H] } 

\def\Teff{$T_{\rm eff}$}
\def\logg{$\log g$}

\newcommand{\tempTeff}{$T_{\rm eff}^{\rm tpl}$}
\newcommand{\templogg}{$\log g^{\rm tpl}$}
\newcommand{\tempFeH}{[Fe/H]$^{\rm tpl}$}

\newcommand\gaia{\textit{Gaia}}
\newcommand\gdr{\rm{DR2} }
\newcommand\kms{\ensuremath{\text{km~s}^{-1}}}
\newcommand\ms{\ensuremath{\text{m~s}^{-1}}}
\newcommand{\epix}{$\mathrm{e}^{-}\mathrm{pix}^{-1}$}

\usepackage{wrapfig}
\usepackage{longtable}
\usepackage{rotating}
\usepackage{lscape}
\usepackage{amsmath}
\usepackage{adjustbox}
\usepackage{subcaption} 
\usepackage{multirow}

\usepackage{graphicx}
\usepackage{txfonts}
\usepackage{todonotes}
\usepackage{pdflscape}
\usepackage{rotating}
\usepackage[font=small,labelfont=bf]{caption}

\usepackage{color}

\begin{document} 

\newcommand\vtidy{\vspace{-\baselineskip}}
\newcommand\etc{{\it etc.\ }}
\newcommand\eg{e.g.\ }
\newcommand\ie{i.e.\ }

\title{{\gaia} Data Release 2}

\subtitle{Processing the spectroscopic data}
\author{
 P.        ~Sartoretti                         \inst{\ref{inst:0001}}
\and D.      ~Katz                            \inst{\ref{inst:0001}}    
\and M.      ~Cropper                       \inst{\ref{inst:0002}}  
\and P.       ~Panuzzo                     \inst{\ref{inst:0001}} 
\and G.M.    ~Seabroke                   \inst{\ref{inst:0002}}
\and Y.       ~Viala                          \inst{\ref{inst:0001}}
\and K.      ~Benson                       \inst{\ref{inst:0002}}  
\and R.      ~Blomme                    \inst{\ref{inst:0003}}
\and G.      ~Jasniewicz                \inst{\ref{inst:0004}}
\and A.       ~Jean-Antoine            \inst{\ref{inst:0005}}
\and H.      ~Huckle                      \inst{\ref{inst:0002}}  
\and M.      ~Smith                       \inst{\ref{inst:0002}}
\and S.        ~Baker                     \inst{\ref{inst:0002}} 
\and F.         ~Crifo                       \inst{\ref{inst:0001}}  
\and Y.         ~Damerdji                \inst{\ref{inst:0006},\ref{inst:0007}}
\and M.        ~David                         \inst{\ref{inst:0008}}
\and C.        ~Dolding                  \inst{\ref{inst:0002}}
\and Y.         ~Fr\'{e}mat                \inst{\ref{inst:0003}}\relax
\and E.         ~Gosset                 \inst{\ref{inst:0007},\ref{inst:0009}}\relax
\and A.         ~Guerrier                 \inst{\ref{inst:thales}}
\and L.P.      ~Guy                           \inst{\ref{inst:0010}}\relax
\and R.        ~Haigron                       \inst{\ref{inst:0001}}\relax
\and K.        ~Jan{\ss}en                 \inst{\ref{inst:aip}}\relax
\and O.       ~Marchal                     \inst{\ref{inst:0001}}
\and G.       ~Plum                      \inst{\ref{inst:0001}}
\and C.       ~Soubiran                 \inst{\ref{inst:0011}}
\and F.        ~Th\'{e}venin            \inst{\ref{inst:0012}}
\and M.       ~Ajaj                         \inst{\ref{inst:0001}}  
\and C.        ~Allende Prieto          \inst{\ref{inst:0002},\ref{inst:tenerife1},\ref{inst:tenerife2}}\relax
\and C.       ~Babusiaux               \inst{\ref{inst:0001},\ref{inst:ipag}}    
\and S.       ~Boudreault               \inst{\ref{inst:0002},\ref{inst:maxplank}}
\and L.        ~Chemin                    \inst{\ref{inst:0011},\ref{inst:0014}}\relax
\and C.        ~Delle Luche                   \inst{\ref{inst:0001},\ref{inst:thales}}
\and C.       ~Fabre                        \inst{\ref{inst:atos}}
\and A.       ~Gueguen                 \inst{\ref{inst:0001},\ref{inst:maxplank1}} 
\and N.C.    ~Hambly                     \inst{\ref{inst:0015}}\relax
\and Y.        ~Lasne                      \inst{\ref{inst:thales}}
\and F.        ~Meynadier               \inst{\ref{inst:0001},\ref{inst:syrte}} 
\and F.        ~Pailler                       \inst{\ref{inst:0005}}
\and C.       ~Panem                     \inst{\ref{inst:0005}}
\and F.       ~Riclet                       \inst{\ref{inst:0005}}
\and F.       ~Royer                       \inst{\ref{inst:0001}}  
\and G.       ~Tauran                    \inst{\ref{inst:0005}}
\and C.       ~Zurbach                   \inst{\ref{inst:0004}}\relax
\and T.       ~Zwitter                       \inst{\ref{inst:0016}}\relax 
\and F.       ~Arenou                \inst{\ref{inst:0001}} 
\and A.       ~Gomez                \inst{\ref{inst:0001}} 
\and V.         ~Lemaitre                    \inst{\ref{inst:thales}}
\and N.       ~Leclerc                        \inst{\ref{inst:0001}} 
\and T.       ~Morel                      \inst{\ref{inst:0007}}
\and U.       ~Munari                 \inst{\ref{inst:asiago}}
\and C.         ~Turon                        \inst{\ref{inst:0001}} 
\and M.        ~\v{Z}erjal                    \inst{\ref{inst:0016},\ref{inst:maruska}}
}     
 \institute{
 GEPI, Observatoire de Paris, Universit\'{e} PSL, CNRS, 5 Place Jules Janssen, F-92190 Meudon, France\relax    
 \label{inst:0001}
  \and Mullard Space Science Laboratory, University College London, Holmbury St Mary, Dorking, Surrey RH5 6NT, United Kingdom\relax                                                                            \label{inst:0002}
\and Royal Observatory of Belgium, Ringlaan 3, B-1180 Brussels, Belgium\relax                                                                                                                                  \label{inst:0003}
\and Laboratoire Univers et Particules de Montpellier, Universit\'{e} Montpellier, CNRS, Place Eug\`{e}ne Bataillon, CC72, F-34095 Montpellier Cedex 05, France\relax                                               
 \label{inst:0004}
\and CNES Centre Spatial de Toulouse, 18 avenue Edouard Belin, F-31401 Toulouse Cedex 9, France\relax                                                                                                          \label{inst:0005}
\and CRAAG - Centre de Recherche en Astronomie, Astrophysique et G\'{e}ophysique, Route de l'Observatoire, Bp 63 Bouzareah, DZ-16340, Alger, Alg\'{e}rie\relax                                                      
 \label{inst:0006}
 \and Institut d'Astrophysique et de G\'{e}ophysique, Universit\'{e} de Li\`{e}ge, 19c, All\'{e}e du 6 Ao\^{u}t, B-4000 Li\`{e}ge, Belgium\relax   
\label{inst:0007}
 \and Universiteit Antwerpen, Onderzoeksgroep Toegepaste Wiskunde, Middelheimlaan 1, B-2020 Antwerpen, Belgium\relax                                                                                            \label{inst:0008}
\and F.R.S.-FNRS, Rue d'Egmont 5, B-1000 Brussels, Belgium\relax                                                                                                                                               \label{inst:0009}
\and Thales Services for CNES Centre Spatial de Toulouse, 18 avenue Edouard Belin, F-31401 Toulouse Cedex 9, France\relax    
\label{inst:thales}  
\and Department of Astronomy, University of Geneva, Chemin d'Ecogia 16, CH-1290 Versoix, Switzerland\relax                                                                                                   \label{inst:0010}
\and Leibniz Institute for Astrophysics Potsdam (AIP), An der Sternwarte 16, D-14482 Potsdam, Germany\relax                                                                                                    \label{inst:aip}
\and Laboratoire d'astrophysique de Bordeaux, Universit\'{e} de Bordeaux, CNRS, B18N, all{\'e}e Geoffroy Saint-Hilaire, F-33615 Pessac, France\relax                                                           
\label{inst:0011}
 \and Laboratoire Lagrange, Universit\'{e} Nice Sophia-Antipolis, Observatoire de la C\^{o}te d'Azur, CNRS, F-34229, F-06304 Nice Cedex, France\relax                                                       
 \label{inst:0012}
 \and Instituto de Astrof\'{\i}sica de Canarias, E-38205 La Laguna, Tenerife, Spain\relax                                                                                                                     \label{inst:tenerife1}
\and Universidad de La Laguna, Departamento de Astrof\'{\i}sica, E-38206 La Laguna, Tenerife, Spain\relax    
\label{inst:tenerife2}  
\and IPAG, Universit\'e Grenoble Alpes, CNRS, IPAG, F-38000 Grenoble, France
\label{inst:ipag}
\and Max Plank Institute f\"{u}r Sonnensystemforschung, Justus-von-Liebig-Weg 3, D-37077 G\"{o}ttingen, Germany
\label{inst:maxplank}
 \and Unidad de Astronom\'ia, Fac. Cs. B\'asicas, Universidad de Antofagasta, Avda. U. de Antofagasta RCH-02800, Antofagasta, Chile
\label{inst:0014}
\and ATOS for CNES Centre Spatial de Toulouse, 18 avenue Edouard Belin, F-31401 Toulouse Cedex 9, France\relax                                                                                                       \label{inst:atos}
\and Max Planck Institute for Extraterrestrial Physics, High Energy Group, Gie{\ss}enbachstra{\ss}e, D-85741 Garching, Germany 
\label{inst:maxplank1}
\and Institute for Astronomy, University of Edinburgh, Royal Observatory, Blackford Hill, Edinburgh EH9 3HJ, United Kingdom\relax                                                                           
 \label{inst:0015}
\and LNE-SYRTE, Observatoire de Paris, Universit\'e PSL, CNRS, Sorbonne Universit\'es, 61 avenue de l'Observatoire, F-75015 Paris, France\relax
\label{inst:syrte}
\and Faculty of Mathematics and Physics, University of Ljubljana, Jadranska ulica 19, SLO-1000 Ljubljana, Slovenia\relax                                                                                         \label{inst:0016}
\and INAF-National Institute of Astrophysics, Osservatorio Astronomico di Padova, Osservatorio Astronomico, I-36012 Asiago (VI), Italy                                           
\label{inst:asiago}
\and Research School of Astronomy and Astrophysics, Australian National University, Canberra, ACT 2611, Australia
\label{inst:maruska}
}

\date{Received \textbf{February 15, 2018}; accepted \textbf{April 8, 2018}}

\abstract
{The \gaia\  Data Release 2 (\gdr) contains the first release of radial velocities complementing the kinematic data of a sample of about 7 million relatively bright, late-type stars.}
 { 
This paper provides a detailed description of the \gaia\ spectroscopic data processing pipeline, and of the approach adopted to derive the radial velocities presented in \gdr.}
{
The pipeline must perform four main tasks: (i) clean and reduce the spectra observed with the Radial Velocity Spectrometer (RVS); (ii) calibrate the RVS instrument, including wavelength, straylight, line-spread function, bias non-uniformity, and photometric zeropoint; (iii) extract the radial velocities; and (iv) verify the accuracy and precision of the results. The radial velocity of a star is obtained through a fit of the RVS spectrum relative to an appropriate synthetic template spectrum. An additional task of the spectroscopic pipeline was to provide first-order estimates of the stellar atmospheric parameters required to select such template spectra. We describe the pipeline features and present the detailed calibration algorithms and software solutions we used to produce the radial velocities published in \gdr.}
 {
  The spectroscopic processing pipeline produced median radial velocities for \gaia\ stars with narrow-band near-IR magnitude \grvs~$\leq12$ (i.e. brighter than $V\sim13$). Stars identified as double-lined spectroscopic binaries were removed from the pipeline, while variable stars, single-lined, and non-detected double-lined spectroscopic binaries were treated as single stars. The scatter in radial velocity among different observations of a same star, also published in \gaia\ DR2, provides information about radial velocity variability. For the hottest  (\Teff~$\geq 7000$ K) and coolest (\Teff~$\leq 3500$ K) stars, the accuracy and precision of the stellar parameter estimates are not sufficient to allow selection of appropriate templates. The radial velocities obtained for these stars were removed from \gdr. The pipeline also provides a first-order estimate of the performance obtained. The overall accuracy of radial velocity measurements is around $\sim200$-300 \ms, and the overall precision is $\sim 1$ \kms;
it reaches $\sim$ 200 \ms for the brightest stars.}
{}
   

\maketitle

\section{Introduction} 

The \gaia\ mission \citep{DR1-DPACP-18} will provide detailed phase space data for distant stars in the Milky Way galaxy, in addition to astrometry,  including radial velocity for many stars ($\sim 150$ million). For logistical reasons, the \gaia\ Radial Velocity Spectrometer (RVS; \citealt{DR2-DPACP-46}) adopts the wavelength range 845-872 nm, which includes the strong \ion{Ca}{ii} triplet lines and is useful for cross-correlation spectroscopy of Galactic stars of most spectral types, except for the earliest spectral types. This paper documents the reduction process and the method used to convert raw observed stellar spectra into the radial velocities presented in the \gaia\ Data Release 2 (hereafter DR2) \citep{DR2-DPACP-36}.


With its two telescopes, \gaia\ continuously scans the sky and collects data of any source detected by the onboard system. Sources are first observed by the astrometric and photometric instruments, when their magnitude in the RVS instrument is estimated. If a source transiting  through one of the four RVS rows (as depicted in Fig~\ref{fig:focalplane}) is bright enough (\grvs~$<16.2$, see Sect.~\ref{section:grvs} below), and if the onboard limit on the number of data that can be obtained simultaneously (corresponding to a source density of $\sim 35,000$ sources deg$^{-2}$ ) is not already reached, a spectrum of the source is recorded by each of the three CCDs on the row. 

Each source will be observed many times during the nominal 60 months of the mission, the expected number of transits per star in the RVS being on average around 40. In the RVS, starlight is dispersed over $\sim1000$ pixels, while the exposure time on the \gaia\ CCDs is fixed at 4.4 seconds by the scanning requirements. The resulting low signal-to-noise ratio (S/N) per pixel for stars near the limiting magnitude of \grvs~$\sim 16$ implies that the combination of many transit spectra will be necessary over the entire mission lifetime for the radial velocities of these stars to be measured. In fact, intermediate data releases such as DR2 are only preliminary. We note that in addition to RVS data for always fainter stars, each future release will include a complete reprocessing of data from the beginning of the mission, with improved calibrations and algorithms.


The RVS spectra processed for \gdr\ were collected during the first 22 months since the start of nominal operations on 25 July 2014. Only stars brighter than \grvs~$\sim 12$ have been processed for DR2, corresponding to $\sim 5\%$ of the spectra acquired by the RVS in this period. Still, the number of spectra treated exceeds 280 million. It is a major challenge to process such large amounts of data, and  in addition to the reliability and robustness of the scientific processing software, the RVS pipeline has to cope with technical issues, such as computational time, data storage, backup, and I/O management for the scientific software and database access.

In this paper, we describe the data processing behind the radial velocities presented in \gdr and the performance achieved. A companion paper \citep{DR2-DPACP-54}
presents a posteriori checks of the DR2 radial velocities and the dependence of the accuracy and precision of velocity measurements on stellar properties.

\begin{figure*}[t]
\begin{center}
\includegraphics[width=0.9\textwidth]{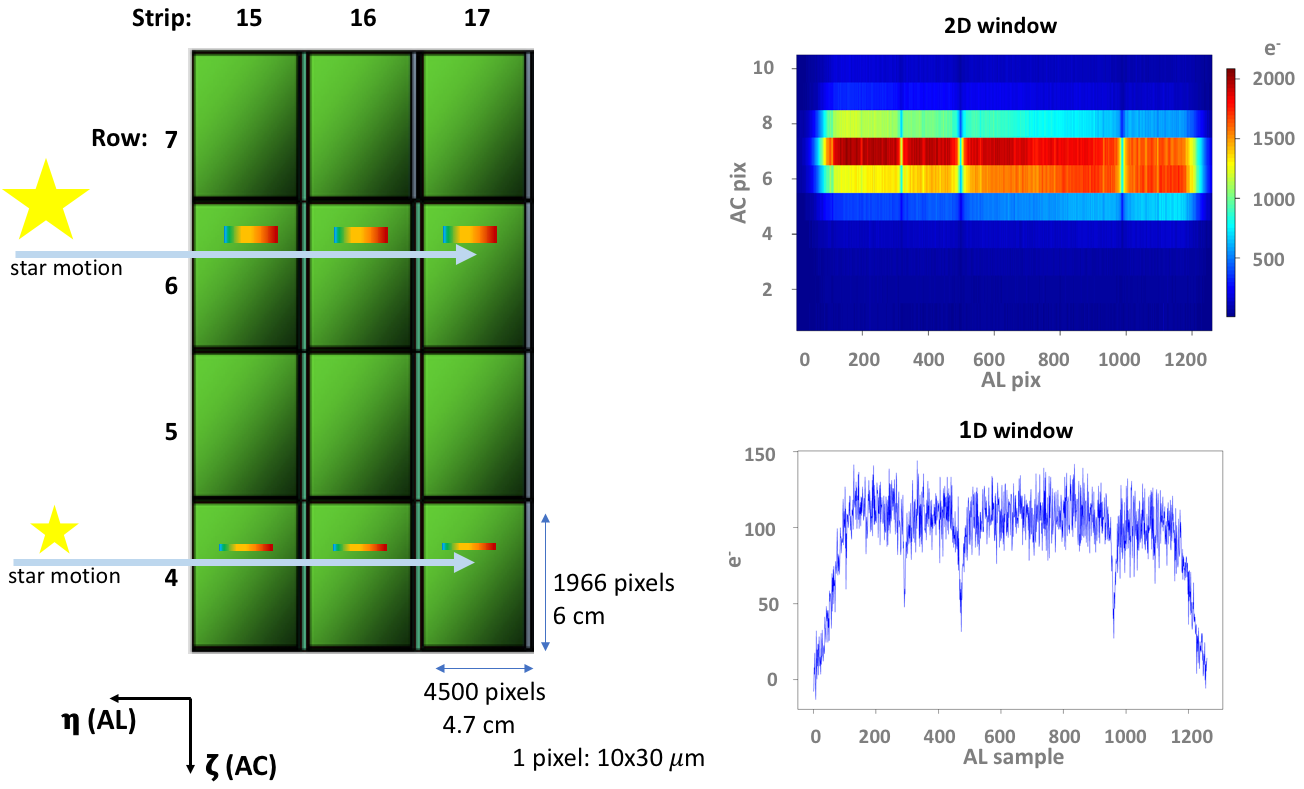}
\caption{RVS focal plane. For the complete \gaia\ focal plane, see \citet[][their Fig. 4]{DR1-DPACP-18}. {\it Left panel:}  12 CCDs of the RVS focal plane laid out in three strips (we use the standard \gaia\ nomenclature and refer to columns as `strips') and four rows. The exposure time is fixed at 4.4 s per CCD (in TDI mode). The star images move in the along-scan (AL) direction (as indicated by horizontal arrows). During each transit, the star crosses all three CCDs on the row, and a spectrum is acquired in each of the three corresponding observational windows centred on the star. If the star has onboard \grvs~$\leq 7$, the telemetered window is 2D, of size 1260- or 1296-AL by 10-AC pixels ({\it upper right panel}). For fainter stars, the 10-AC pixels are summed during read-out to produce a 1D window ({\it lower right panel}). The FoV of both \gaia\ telescopes are projected onto the focal plane. The orientation of the field angles $\eta$ and $\zeta$ is indicated in the lower left corner. The origin of the axes is in the astrometric focal plane and is different for the two FoVs. 
\label{fig:focalplane}}
\end{center}
\end{figure*}

\section{Overview and limitations of the spectroscopic pipeline}\label{sec:overview}

\subsection{ \grvs\ magnitude}
\label{section:grvs}

\grvs\ is a narrow-band near-IR (NIR) magnitude, whose nominal passband (845--872 nm) is described in \cite{2010A&A...523A..48J}. The impracticability of estimating uniformly the \grvs\ magnitude for on-the-fly data acquisition, spectroscopic pipeline selection, and from the flux in the spectrum has led to different definitions of this quantity. 

\begin{itemize}
\item
The onboard \grvs\ magnitude is estimated by the onboard software. It is used to decide whether to allocate an RVS window to a star (stars fainter than \grvs~$=16.2$ do not get an observation window), as well as which window type to allocate (2D if \grvs~$ \leq 7$, 1D otherwise; see Fig. \ref{fig:focalplane}). The onboard \grvs\ magnitude is estimated using the flux in the portion of the red photometer (RP) spectrum covering the RVS wavelength range, or, when the RP spectrum is saturated, the flux in the astrometric image. The \grvs\ magnitude is derived in each of these two cases using had hoc formulae whose parameters were calibrated during commissioning. 

\item 
The onboard \grvs\ magnitude described above is contaminated by straylight and instrumental effects. Another estimate of the \grvs\ magnitude was therefore used to select the stars to process through the DR2 spectroscopic pipeline: the external \grvs\ magnitude, noted \extgrvs. This is taken to be the magnitude listed in the initial Gaia source list (IGSL) published before the mission started \citep{2014A&A...570A..87S}. For the 8\% of stars detected by \gaia\ not found in the original IGSL, we take the \extgrvs\ magnitude to be the onboard one. Stars brighter than \extgrvs~$=12$ are processed through the DR2 spectroscopic pipeline, the brightest ones (\extgrvs $\leq 9$) being used for wavelength calibration. For reference, for most ($\sim$ 85\%) stars in the IGSL, the \extgrvs\ magnitude was estimated from Tycho-2 $B_T$ and $V_T$ and GSC23 $B_J$ and $R_F$ \citep{2008AJ....136..735L} magnitudes using the formulae \vspace*{0.3cm}\\
\extgrvs $ = R_F -0.0974   -0.4830(B_J-R_F)   -0.0184(B_J-R_F)^2-0.0178(B_J-R_F)^3$ \\       
\extgrvs $ = V_T - 0.1313 - 1.3422(B_T-V_T)    - 0.09316(B_T-V_T)^2 - 0.0663(B_T-V_T)^3$\,. \vspace*{0.3cm}\\  
This estimate was revealed to be imprecise, especially in case of red colour. A better estimate is obtained using $V$ and $I$ (see Sect.~\ref{sssec:auxgrvs}).

\item
Another quantity is computed internally by the pipeline directly from the spectra. This internal \grvs\ magnitude, noted \intgrvs, is computed based on the flux integrated in the RVS spectrum and a calibrated reference magnitude (\refgrvs, see Sect.~\ref{sssec:grvszp}) relying on $V$ and $I$ observations of about 113\,000 {\it Hipparcos} stars (see Sect.~\ref{sssec:auxgrvs} and \ref{sssec:intmag}).

\end{itemize}

\subsection{Purpose of the spectroscopic pipeline}

The radial velocity is estimated by measuring the Doppler shift of the spectral lines in the observed RVS spectrum compared to a synthetic template spectrum (at rest), selected to be as similar as possible to the observed spectrum. 

The first goal of the \gaia\ spectroscopic pipeline is to measure the all-transit-combined radial velocity (\RV) for all the \gaia\ stars observed by the RVS (i.e. brighter than onboard \grvs~$\sim16$), for which an appropriate template could be found (the atmospheric parameters of the stars needed to find the template will be available from another pipeline). 
The precision depends on the spectral type, the magnitude of the star, and the number of observations. The pre-launch end-of-mission requirements on the precision of the radial velocity measurements (mission averaged, 40 transits) were set on three representative spectral types, B1V, G2V, and K1IIIMP (where MP stands for metal-poor, i.e. \FeH~$\approx -1.5$ dex), depending on the magnitude,
\begin{itemize} 
\item 1 \kms: B1V with \grvs~$\leq 7.2$ ($V \sim 7$); G2V with \grvs~$\leq 12.1$ ($V \sim 13$) and  K1IIIMP with \grvs~$\leq 12.3$ ($V \sim 13.5$);
\item 15 \kms: B1V with \grvs~$\sim 12.2$ ($V \sim  12$); G2V with \grvs~$\sim 15.6 $ ($V \sim 16.5$) and K1IIIMP with \grvs~$\sim 15.8$ ($V \sim 17$).
\end{itemize}
The precision on the \RV\ measurements is expected to be low for early-type stars, which in the narrow NIR RVS wavelength range have shallow and broad spectral lines, and for late M-type stars, dominated by the molecular TiO band. For medium-temperature stars (FGK types), the precision is expected to increase towards cooler stars, where more neutral lines and more contrast with the continuum provide more information for comparisons with templates. Some RVS spectra of stars with different spectral types are shown in \citet[][their Fig. 17]{DR2-DPACP-46}. 
The end-of-mission performance is modelled  for several spectral types in \citet[][their Sect. 8.3; see also \url{https://www.cosmos.esa.int/web/gaia/rvsperformance}]{DR1-DPACP-18}. With regard to systematic uncertainties, 
the pre-launch end-of-mission requirement was that after calibration, these should be less than $300$~\ms \citep[][Table 1]{DR2-DPACP-46}.

Another goal of the spectroscopic pipeline is to provide the radial velocity at each transit through the RVS (\RVtr) for the brightest stars to identify variable candidates, detect double-lined spectra, and estimate the velocity of the two components. The single-transit radial velocities are used by the downstream pipelines for variability and multiple-system studies and will be published in DR4.  Rotational velocities will be estimated for a subset of bright stars.  
The all-transit-combined spectra are also produced and used by downstream pipelines to estimate stellar atmospheric parameters and abundances \citep{2016A&A...585A..93R}. A subset of the combined spectra will be published in DR3.

The specific goal for DR2, with 22 months of mission data, was
to provide the all-available-transit-combined (a minimum number of two transits is required) \RV\ for all the stars 
observed by the RVS brighter than \extgrvs $= 12$, for which an appropriate template could be found. The aim was to approach the end-of-mission requirements as best possible, and reach a precision close to $\sim$1 \kms. The single-transit radial velocities, \RVtr\ and the combined spectra are also obtained, but not published.
One problem in the DR2 pipeline was that estimates of the atmospheric parameters of observed stars, which are needed to associate{} synthetic templates, were not available for most stars. As a result, an additional goal of the pipeline was to provide (rough) estimates of these stellar parameters, to be able to associate appropriate template. For the hottest  (\Teff~$\geq 7000$ K) and coolest (\Teff~$\leq 3500$ K) stars, however, the accuracy and precision of such estimates are not sufficient to allow selecting appropriate templates, and the radial velocities obtained for these stars have been removed from DR2. Moreover, no template is available for emission-line stars, which were removed from DR2. 

Further goals of the pipeline are to automatically check whether the accuracy and precision of the computed radial velocities agree with expectations, and to self-check the behaviour of each processing step to detect any potential problem.

\subsection{Pipeline overview}

\begin{figure*}
\begin{center}
\includegraphics[width=0.8\textwidth]{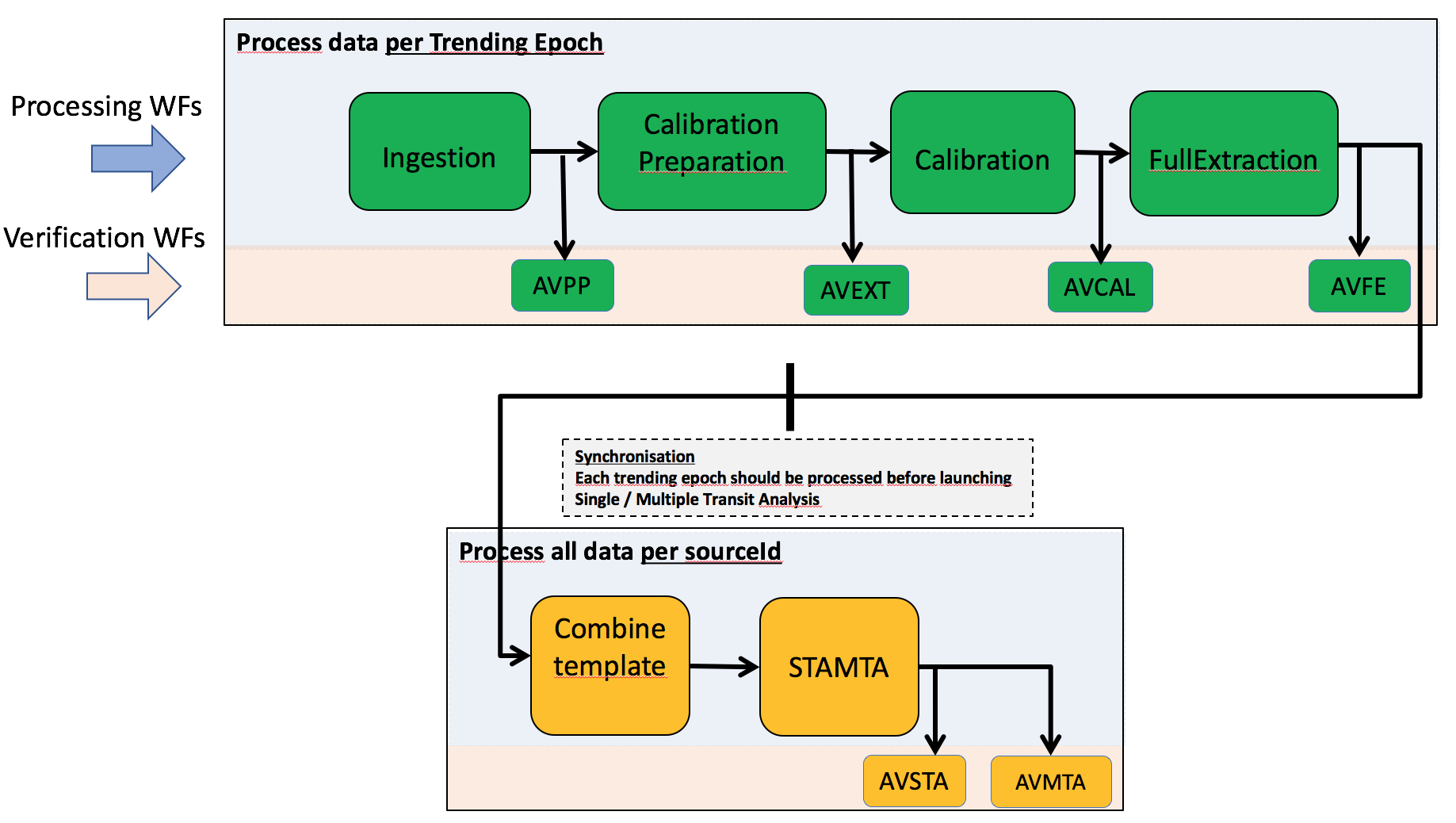}
\caption{Flow chart of the spectroscopic pipeline. The pipeline is composed of six processing workflows and six verification workflows. The green workflows process the data per transit (i.e. per observation). The yellow workflows process the data per source and collect the information produced upstream from each transit.
\label{fig:workflows}}
\end{center}
\end{figure*}

Figure \ref{fig:workflows} shows a flowchart illustrating the spectroscopic pipeline. We provide here an overview of the primary goals of each workflow (step) of this pipeline, which are described in detail in the following sections.

First, the {\tt Ingestion} workflow extracts information from the input data (Sect. \ref{sec:inputdata}) and auxiliary data (Sect.~\ref {sec:auxdata}) and groups this information into data packets that are processed by the downstream workflows: the position of the source and its coordinates in the focal plane (Fig.~\ref{fig:focalplane})
are recorded, which are required for wavelength calibration; overlapping spectra are removed; the sourceId to which the spectrum belongs is identified and searched for in the IGSL to determine the associated \extgrvs. Stars fainter than  \extgrvs~$ = 12$ are removed from the pipeline; the sourceId is also searched for in the auxiliary catalogues to identify standard stars (Sect.~\ref{sssec:auxstd}) and store their reference radial velocities, which are used in the {\tt Calibration} workflow to fix the wavelength calibration zeropoint; the sourceId is also used to identify stars with atmospheric parameters available from the literature (Sect. \ref{sssec:auxparams}), to which the appropriate synthetic spectrum is then associated (Sect.~\ref{sssec:synthspe}), and to identify stars with available \refgrvs\ (Sect. \ref{sssec:auxgrvs}) to be used as reference for the \intgrvs\ zeropoint calibration.

In the {\tt Calibration Preparation} workflow (Sect. \ref{sec:extraction}), stars brighter than \extgrvs~$= 9$ are selected to be used as calibrator stars, and their spectra are prepared to be used in the {\tt Calibration} workflow; the spectra are corrected for electronic bias and dark current; the gain is applied; the straylight background and cosmic rays are removed; the 2D windows are collapsed into 1D; a first-guess model of the wavelength calibration is applied; the atmospheric parameters (if not available from the literature) are estimated; stars with $4500 < $~\Teff$~< 6500$ K are selected as calibrator stars for the wavelength calibration, together with standard stars; for the \intgrvs\ zeropoint calibration, only stars with available \refgrvs\ are selected.

The {\tt Calibration} workflow (Sect. \ref{sec:calibration}) processes stars prepared for calibration (i.e. with \extgrvs~$\leq9$): the wavelength-calibration model is computed (Sect.~\ref{sssec:wavecalib}); stars with $4500 < ~$\Teff$~< 6500$ K, along with radial-velocity standard stars, are used to fix the wavelength calibration zeropoint; the \grvs-zeropoint calibration model is also computed (Sect. \ref{sssec:grvszp}), using stars with available \refgrvs\ as photometric standards. The {\tt Calibration}  workflow also incorporates calibration parameters estimated off-line (as described in Sect.~\ref{sec:calibration}), to be applied to the spectra in the downstream workflows.

In the {\tt FullExtraction} workflow, the spectra of all stars (including those with  $9<$~\extgrvs~$ \leq 12$) are treated as in the {\tt Calibration Preparation} workflow above, but this time using the calibration parameters obtained in the {\tt Calibration} workflow. The stellar atmospheric parameters are determined, and a synthetic template is associated with each transit of a star. 

In the {\tt Combine Template} workflow (Sect.  \ref{sssec:combinetemp}), the templates associated with different transits of a same star are combined to ensure that a single template per star is used to determine its radial velocity.

In the {\tt STAMTA} workflow, the radial velocity of a star is computed separately for each transit, \RVtr\ (Sect. \ref{ssec:sta}). Then, the median of the single-transit radial velocities obtained from all available transits is computed (Sect. \ref{ssec:mta}) to obtain the final radial velocity of the star, \RV, published in DR2. 

The purpose of the {\tt Automated Verification} workflows ({\it i.e.} {\tt AVPP, AVEXT, AVFE and AVSTA and AVMTA} in Fig. \ref{fig:workflows}) is to check the products to be used by the downstream workflows: the data required for the verification diagnostics are processed in the pipeline to obtain the parameters to monitor; then, they are sent to an off-line software that produces diagnostic plots made available for validation via a web interface.

The entire process is managed by a software system, named SAGA, and the code is run in parallel on an Hadoop cluster system with 1100 cores and 7.5 TB memory. The processing took the equivalent of 630\,000 hours CPU time and needed 290 TB disc space. 

\subsection{Data products in \gdr}
The products of the spectroscopic pipeline published in {\it Gaia} DR2, the notation used in this paper, and their name in the \gaia\ Archive (\url{https://gea.esac.esa.int/archive/}) are listed below.
\begin{itemize}

\item
Median radial velocity (\kms): noted \RV; \gaia\ Archive name: {\it radial\_velocity}. This is provided for most of the stars with $4 \leq G_{\rm{RVS}} \leq 12$. No variability detection is attempted, and all stars are treated as single stars. The detected SB2s (Sect. \ref{sssec:cu6spe_TodCorLight}) and emission-line stars are removed from DR2.
\item
Radial velocity uncertainty (\kms): noted \RVerr; \gaia\ Archive name: {\it radial\_velocity\_error}.
\item
Number of transits: the number of transits used to determine \RV, noted $N$; \gaia\ Archive name:  {\it rv\_nb\_transits}.
\item
Template temperature (K): \Teff\ of the template used to determine \RV, noted \tempTeff; \gaia\ Archive name: {\it rv\_template\_teff}. 
\item
Template surface gravity (dex): \logg\ of the template used to determine \RV, noted \templogg; \gaia\ Archive name: {\it rv\_template\_logg}.
\item
Template metallicity (dex): \FeH of the template used to determine \RV, noted \tempFeH ; \gaia\ Archive name: {\it rv\_template\_fe\_h}.
\end{itemize}

\subsection{Limitations of the DR2 pipeline}
As the other \gaia\ pipelines, the spectroscopic processing system will evolve and will be upgraded with new functionalities in the future data releases. In \gaia\ DR2, the radial velocities of stars brighter than \grvs\ $\leq12$ are provided, and the pipeline has the first-order functionality sufficient to treat these bright stars. The final \RV\ provided for each source is the median value of the radial velocities estimated from all source transit spectra. 

In \gaia\ DR3, the expectation is to provide \RV\ for stars down to a magnitude \grvs~$\sim14$. Then, second-order calibrations and the functionality to de-blend spectra will be included in the pipeline. The final \RV\ for each star will be computed with a more reliable model combining information from all single-transit spectra of the star. 

In addition to the limitations from still-missing functionalities, the DR2 version of the RVS pipeline does not have access to the effective temperature (\Teff)~ published in \gdr \citep{DR2-DPACP-43}, which was produced by other processing pipelines running at the same time, nor to the $G$, $G_{BP}$ and $G_{RP}$ magnitudes (and the associated refined estimates of the \grvs\ magnitude) from \citet[][Eq. 2 ]{DR2-DPACP-36}.  As mentioned above, the \Teff\ needed to associate a spectral template with a star is taken from the literature when available, and otherwise estimated by the pipeline (Sect. \ref{sssec:detAP}). The \extgrvs\ magnitude used to select the stars to be processed is computed as described in Sect.~\ref{section:grvs}.

Despite these limitations, the overall accuracy and precision of the radial velocities measured by the pipeline and published in DR2 are close to end-of-mission requirements (Sect.~\ref{sec:results}).

\section{Input data \label{sec:inputdata}}

The time on board \gaia , the OBMT (onboard mission timeline), generated by the \gaia\ onboard clock, counts the number of six-hour spacecraft revolutions since launch. The events on board are given in OBMT (the notations REV, revolutions, are also used). The relation to convert OBMT into barycentric coordinate time (TCB) is provided by Eq. 1 in \cite{DR2-DPACP-36}. 

The spectra processed by the DR2 pipeline have been acquired by the RVS between OBMT 1078.3795 (25 July 2014) and OBMT 3750.5602 (23 May 2016). 

In the focal plane (see Fig.~\ref{fig:focalplane}), two directions are defined relative to the scan direction: AL (along scan) and AC (across scan). In the spectra, AL is the dispersion direction and AC is the spatial direction. The spectra are acquired with 1260 or 1296 AL pixels depending on the onboard software version used; to mitigate the straylight effect, an updated onboard software, VPU version 2.8 \cite[see][]{DR2-DPACP-46}, was put in operation in June 2015, and the spectra acquired since then have 1296 samples. The wavelength range is 845-872 nm and the resolution element is $\sim 3$ AL pixels. The RVS instrument and the spectra produced are described in \citet{DR2-DPACP-46}, and the in-flight performance of the \gaia\ CCDs in \citet{DR1-DPACP-20}.

To be treated by the spectroscopic pipeline, the telemetered RVS spectra are reformatted by the initial data treatment (IDT) pipeline \citep{DR1-DPACP-7}, and so is the associated information necessary for processing the spectra, such as the detection features: time, coordinate, field of view (FoV), CCD row, solar rotation phase, onboard magnitude, the AC position of the window on the CCD, its size and truncation status (the windows, originally of AC=10 pixel size, can be truncated if in conflict with other windows), and the pre-scan pixel values necessary to estimate the electronic bias. The main input data to the spectroscopic pipeline are the IDT products called {\it SpectroObservation}, containing the three CCD spectra corresponding to one transit (i.e. one single observation) of the source. The IDT also provides the files {\it BaryVeloCorr} containing the value of the barycentric velocity correction calculated every 5 minutes. The barycentric velocity correction is added to the spectroscopic, {\it Gaia}-centric, radial velocity measurements to obtain the radial velocities relative to the solar system barycentre.

In addition to the IDT, the spectroscopic pipeline depends on the data produced by the intermediate data update (IDU). The IDU cross-match permits source identifier (sourceId) to be associated to each SpectroObservation. The star coordinates from the astrometric global iterative solution (AGIS) and the spacecraft attitude 
permit calculating the field angles ($\eta, \zeta$), i.e. the coordinates in the FoV reference system needed by the wavelength calibration. For information on the AGIS2.1 solution used in the spectroscopic pipeline, see \citet{DR2-DPACP-51}, and for a description of the field angles and of the FoV reference system, see \citet{2012A&A...538A..78L} and \citet{DR1-DPACP-7}. 

The IGSL is also part of the inputs to obtain the information on the \extgrvs\ of the star.

\subsection{Extraction of the input data information}\label{sssec:ingestion}

In the {\tt Ingestion} step (Fig. \ref{fig:workflows}), the information contained in the input data that is relevant for the downstream processing is extracted and stored in the format needed by the spectroscopic pipeline data model. During this step, the data to be processed downstream are selected (see Sect. \ref{sssec:inputselection}). 

For each CCD spectrum, some additional information is computed using the IDT and IDU input data:
\begin{itemize}
\item The time $t^{\rm obs}$ is calculated for each of the three CCD windows in a transit. This is the time at the mid point of the window as it passes the CCD fiducial line. The fiducial line is the nominal mean position of those light sensitive
TDI lines (one CCD AL pixel corresponds to one TDI and to 982.8 $\mu$s) that contribute to the signal integration. 
Based on the time, $t^{\rm obs}$, of the mid point of the window, the time  $t_s^{\rm obs}$ of the samples of the window are calculated.

\item The field angles, $\eta(t_s^{\rm obs})$ and $\zeta(t_s^{\rm obs})$, are calculated for each sample of the spectrum using the AGIS astrometric parameters
$\alpha$, $\delta$, $\mu_\alpha$, $\mu_\delta$, $\varpi$, and the attitude OGA3 \citep{DR2-DPACP-51}.  
 
 \item The barycentric velocity correction is calculated at any transit time by linear interpolation of the {\it BaryVeloCorr} data provided by the IDT.
 This correction is added to the measured {\it Gaia}-centric radial velocity (Sect. \ref{ssec:sta}) in order to obtain a barycentric radial velocity.

\end{itemize}

\subsection{Selections applied to the Input data}\label{sssec:inputselection}

The spectra that could not be processed by this version of the pipeline were excluded from the data flow. The result is that the large majority (more than 95\%) of the spectra acquired by the RVS are excluded in DR2, but are awaiting inclusion in future releases. The following criteria were used to filter out the data:

\begin{itemize}
\item Spectra of sources fainter than \extgrvs=12 are filtered out. This is the most important filter, as the large majority of the RVS data are fainter than 12 (the faintest RVS spectra have onboard \grvs $=16.2$). The value of \extgrvs=12 was chosen based on general considerations: it corresponds to a $S/N \sim 2$ in a single CCD spectrum, assuming the median straylight level observed at the beginning of the spacecraft operations ($\sim$ 5 $e^- pix^{-1}$). 

However, the \extgrvs \  is only a rough estimation, and many fainter stars (based on their flux observed in the window and their \intgrvs) entered the processing despite this filter.

\item Spectra with non-rectangular truncated windows are filtered out. This filter is at transit level: depending on the particular observation geometry during a transit, a source spectrum may
or may not overlap with one or more other source spectra. This filter excludes approximately 40\% of the remaining spectra. The overlapping is more important in crowded regions, where 60\%-80\% of the spectra have a truncated window. 

Because some bright stars produce nearby spurious sources, they have a truncated but rectangular window, and are not filtered out. Their spectra exhibit a flux loss, which combined with non-perfect AC centring of the window and the tilt of the spectrum can result in an artificial spectral slope. Approximately 40\% of the stars with onboard magnitude \grvs\ between 7 and 9 are affected and have a window AC size of 5 pixels (instead of the normal size of 10). The spurious detection events decrease rapidly with the magnitude of the star, and at \grvs\ $\sim 11$, only $\sim$ 5\% of the stars have a reduced size AC window. The 2D windows of the very bright stars are never truncated because of other source windows, and may be truncated only by the CCD borders.

\item Sources without AGIS coordinates are filtered out. So are spectra acquired during pre-defined time intervals, where the data are known to be of poor quality. These intervals include the time during the decontaminations, the refocusing, the commissioning of the VPU 2.8, the time intervals where the AGIS residuals are high, and the time gaps in the IDT barycentric velocity correction that are longer than 10 minutes. 

The total time covered by all the excluded intervals is $\sim$ 200 revolutions or $\sim$ 7.5\% of the total observation time.

\end{itemize}

\section{Auxiliary data\label{sec:auxdata}}

In this section, the auxiliary non-\gaia\ data used by the pipeline are described. Some of them play an important role in the calibration and in the determination of the \RV\ results.

\subsection{Auxiliary radial velocities of standard stars}\label{sssec:auxstd}

The radial velocity standard stars used by the wavelength-calibration modules have \RV\ that is constant in time and precise at the level of 0.1 \kms, as determined from high-precision ground-based observations. They are needed as external calibrators to fix the RVS \RV\ zeropoint. 
The auxiliary \RV\ catalogue, including the 2568 standard stars with the highest precision and accuracy used in wavelength calibration, is described in \cite{DR2-DPACP-48}. The \RV\ zeropoint of that dataset is set by the SOPHIE spectrograph  \citep{2008SPIE.7014E..0JP}. By being calibrated on this dataset, the RVS \RV\ zeropoint should agree with that in \cite{DR2-DPACP-48}.

The wavelength calibration software module, which works per CCD and FoV (Sect. \ref{sssec:wavecalib}), requires standard stars covering the AC dimension of each CCD in order to produce good results in each calibration unit (a calibration unit consists of a dataset covering 30 hours of observations and containing the calibrator stars; see also Sect. \ref{sssec:wavecalib}). This is achieved, in general, by observing approximately 200-300 standard stars every 30 hours. 

An additional 5729 stars were used as standard stars. They were extracted from the master list 
in \cite{2010A&A...524A..10C}, where the photometric variables and known multiple-systems are excluded, and only the stars with F-G-K spectral types were kept. The list was then correlated with the 
Extended Hipparcos Compilation (XHip) \citep{2012AstL...38..331A}, and only stars with a \RV\ measurement with an uncertainty 
\RVerr $ < 1$ \kms\ and quality A and B were retained.  
Finally, the list of candidate standards was refined using the RVS observations obtained during the trial runs of the pipeline, together with the observations obtained by the daily pipeline. The stars with an RVS \RV\ different by more than 3 \kms\ from the one in XHip and the stars that varied between the RVS single-transit measurements $>$ 3 \kms\ were excluded from the list. We also removed stars with a double-lined spectrum. 

Table \ref{tab:auxradvel} lists the standard star catalogue CU6GB-cal (derived from the name of the \gaia\ spectroscopic team; GB stands for ground-based) and XHip. Figure \ref{fig:auxradvel-grvs} shows the magnitude \extgrvs\ distribution of the stars belonging to these catalogues.

\subsection{Auxiliary radial velocities of validation stars}\label{sssec:auxval}
\begin{figure}[h]
\includegraphics[width=0.5\textwidth]{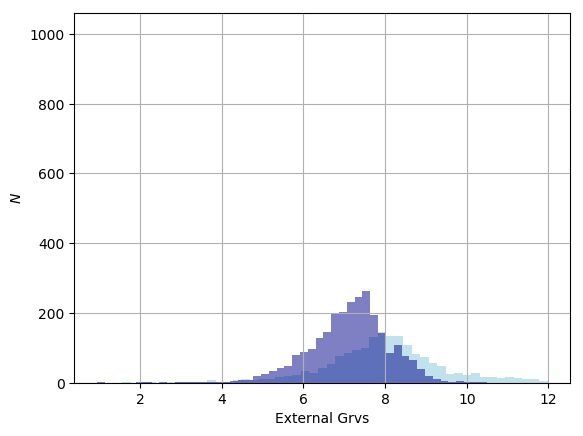}
\includegraphics[width=0.5\textwidth]{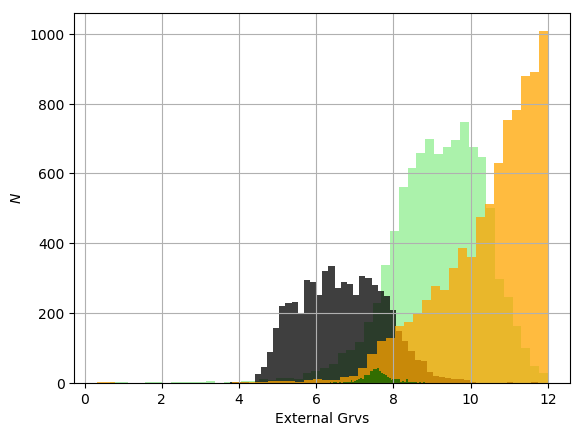}
\caption{Distribution on \extgrvs\ of the stars belonging to the auxiliary radial velocity catalogues that are processed by the pipeline. {\it Top:} Magnitude distribution of the stars in CU6GB-cal (dark blue) and in CU6GB-val (light-blue). {\it Bottom}: Magnitude distribution of the stars in XHip (black), RAVE (light green), APOGEE (orange), and SIM (dark green). }
\label{fig:auxradvel-grvs}
\end{figure}

The validation dataset consists of stable stars for which a ground-based radial velocity is available. Table \ref{tab:auxradvel} also lists the catalogues used for validation in the automated verification workflow (Sects. \ref{sssec:avsta} and \ref{sec:results}) and in the off-line validation \citep{DR2-DPACP-54}. Figure~\ref{fig:auxradvel-grvs} shows the \extgrvs\ magnitude distribution of the stars processed by the pipeline. The selection of the stable stars was based on the following criteria: 
\begin{itemize}
\item CU6GB-val: All the stars from \cite{DR2-DPACP-48} that
were not used as CU6GB-cal were used as validation stars. These stars are presumed to be stable based on at least two high-precision ground-based observations over a minimum time baseline of 30 days and with a standard deviation $< 0.3$ \kms. It includes observations with standard deviation $< 0.1$ that do not fulfil all the RVS requirements for calibrations such as magnitude and spectral type, and some new observations qualified for calibration that could not be integrated in the pipeline processing workflows.
\item RAVE: Stars from DR5 with a constant radial velocity within 1.5 \kms during at least two RAVE observations \citep{LL:TZ-005}.
\item APOGEE: Stars from DR3 that exhibit a constant radial velocity within 0.5 \kms (i.e. the scatter on the radial velocity, \textit{vscatter} $\leq 0.5$ \kms) during at least four APOGEE observations, and the $\chi^2$ probability for non-constancy is $<$ 0.1 (\textit{stablerv\_chi2\_prob} $<$ 0.1). 
\item SIM: Stars from the Radial Velocities of the Southern Space Interferometry Mission Grid Stars Catalogue \citep{2015MNRAS.446.2055M} have a binarity probability < 95\%. 
\end{itemize}

\begin{table}
\begin{center}
\begin{adjustbox}{width=0.5\textwidth}
\begin{tabular}{l c c c c c }
  \hline\hline\\[-0.3cm]
  Cat Name & nb stars & nb transits& $\sigma_{V_R}$ & pipeline use \\
  &&&\kms&\\
  \hline\\[-0.3cm]
   CU6GB-cal & 2568 &34630 & < 0.1 &  calibration\\
   XHip  & 5729  & 83366 &  < 1.0 & calibration\\
  \hline\\[-0.3cm]
  CU6GB-val & 1819 & 16819 & < 0.3& validation\\
  RAVE &  9542  &  86035& < 1.5& validation\\
  APOGEE  & 8584  &  76177 &< 0.5& validation\\
  SIM   & 650  &  4885 & < 0.1 & validation\\
 \hline
\end{tabular}
 \end{adjustbox}
\caption{External radial velocity catalogues used in the pipeline. The first two catalogues are used in wavelength calibration and are described in Sect. \ref{sssec:auxstd}. The other catalogues (Sect. \ref{sssec:auxval}) are used to verify the RVS pipeline results. $\sigma_{V_R}$ is the uncertainty associated with the radial velocity measurements provided by each catalogue.}
\label{tab:auxradvel}
\end{center}
\end{table}

\subsection{Auxiliary atmospheric parameters}\label{sssec:auxparams}

The atmospheric parameters (\Teff, \logg\ and \FeH) of the observed stars are used for two tasks in the pipeline: \begin{itemize} 
\item to select the synthetic template that matches the stellar spectrum (the template is selected based on the minimum distance between the atmospheric parameters of the star and those of the synthetic spectrum, Sect. \ref{sssec:detAP});
\item to select calibration stars with a  suitable spectrum for the calibration in question. Specifically, the wavelength calibration needs spectra with pronounced \ion{Ca}{ii} lines, therefore the selection is $4500 < T_{\mathrm{eff}} < 6500$ K.
\end{itemize}

Because the \Teff\  measurements obtained with \gaia\ data in \citet{DR2-DPACP-43} were not available to the pipeline, a list of auxiliary atmospheric parameters was compiled. It contains parameters, mostly \Teff, for $\sim $1.8 million stars. 

The auxiliary catalogue, AuxAtmParams, contains $\sim$ 800\,000 values of \Teff\   and $\sim$ 650\,000 values of \logg\ and \FeH\ taken from the literature \citep{LL:CS-011}, and  another $\sim 1$ million values of \Teff\ that have been derived for Tycho2 stars (with $4500 \leq T_{\mathrm{eff}} \leq  7500$ K) using the 2MASS photometry
point-source catalogue (with the quality flags taken into account) and the \cite{2010A&A...512A..54C}
effective temperature versus $J-K$ colour relations [T. Zwitter, internal \gaia\ communication].

Given that some stars appear in several catalogues, a priority selection was performed, following \cite{LL:CS-011}. Priority was given to the spectroscopic over the photometric estimation, and for the photometric catalogues, preference was given to those providing the three parameters instead of only \Teff. 

Only some (i.e. $\sim$ 15\%) of the stars treated in \gdr are in the auxiliary atmospheric parameter files and have set atmospheric parameters. The treatment for the other stars is described in Sect. \ref{sssec:28templates}.

\subsection{Auxiliary synthetic spectra}\label{sssec:synthspe}
Synthetic spectra were used to generate the
template spectra that simulate noiseless RVS spectra (Sect. \ref{sssec:generateTemplate}). The templates were cross-correlated with the RVS spectra both for the \RV\ determination (Sect. \ref{sssec:sta}) and in the wavelength calibration to determine the spectral line positions (Sect. \ref{sssec:wavecalib}).
  
The auxiliary synthetic spectra library used in the pipeline is composed of 5256 spectra selected from an updated version of the synthetic spectral library described in \cite{2011JPhCS.328a2006S}. The  \cite{2011JPhCS.328a2006S} spectral library was updated, and the selection of the grids for the spectroscopic pipeline is described in \cite{LL:RHB-005}. The main improvement was the extension of the MARCS grid \citep{2008A&A...486..951G} to the cooler stars. The following grids were selected:
\begin{itemize}
\item Spectra obtained with MARCS models:
  \begin{itemize}
  \item $T_{\rm eff}$: 2500 - 8000 K;  step 100 K between 2500 - 3900 K and 250 K between 4000 - 8000 K; 
  \item \logg: $-0.5$ to +5, step 0.5; 
  \item \FeH: $-5.0, -4.0, -3.0, -2.5, -2.0, -1.5, -1.0, -0.75$, $-0.50, -0.25, 0.0, +0.25, +0.5, +0.75, +1.0$;
  \end{itemize}
\item A-type spectra:
 \begin{itemize}
  \item $T_{\rm eff}$: 8500 - 15\,000 K;  step 500 K;
  \item \logg: +0.5 to +5, step 0.5; 
  \item \FeH: $-0.5$ to +0.25, step 0.25;
  \end{itemize}
\item OB-type spectra:
\begin{itemize}
\item
\Teff: 15\,000 - 55\,000 K, step 1000 K between 15\,000 - 30\,000 K, step 2500 K between 30\,000 - 50\,000 K;
\item \logg: highest value: +4.75, lowest: approximately linearly from 1.75 at 15\,000K to 4.0 at 55\,000 K;
\item \FeH : $-0.3$, 0.0, +0.3.
\end{itemize}
\end{itemize}

The synthetic spectra are characterised by a number of parameters. The main parameters are effective temperature (\Teff), gravity (\logg),
and metallicity (\FeH). Secondary parameters include the abundance of the $\alpha$ elements ([$\alpha/{\mathrm{Fe}}$]),
and the turbulent velocity ($v_{\rm turb}$). 

The selection was made to reduce the number of spectra 
to those that could be exploited by the pipeline, and to have only one spectrum per set of main parameters. 
The MARCS spectra are available for various values of the [$\alpha$/Fe] parameter, and the
lowest non-negative $[\alpha$/Fe] for a given \Teff, \logg,\  and \FeH\ was selected.
The OB spectra are available with two values of $v_{\rm turb}$ , and those with 
$v_{\rm turb} = 2~{\rm km\,s}^{-1}$ were selected.
The A-star and OB-star grids overlap at \Teff = 15\,000 K. In that case, we selected
the OB-star spectra because that grid is denser. 

The current set of synthetic spectra does not include convective
motions or gravitational redshift. At a later stage, the synthetic spectra are planned to be replaced by versions that
do include convective motions \citep[e.g.][]{2011JPhCS.328a2012C, 2018arXiv180101895C, 2013A&A...550A.103A}.

\subsection{{\bf Synthetic template spectra}}\label{sssec:28templates}

For those stars without atmospheric parameters in the auxiliary catalog (the majority), a pipeline module, called {\tt {\tt DetermineAP}} (Sect. \ref{sssec:detAP}) was used to determine their atmospheric parameters by cross-correlation with the star spectrum. The template that matched the RVS spectrum best was selected, and the atmospheric parameters associated with this template were attributed to the star.

This restricted template spectral library, containing 28 spectra, was created for {\tt DetermineAP}, and contains a subset of the synthetic spectra data (Sect. \ref{sssec:synthspe}).
The following arguments were used to construct this restricted synthetic template data set:
\begin{itemize}
\item The set should be relatively small, so that the computation time is short.
\item The steps in the atmospheric parameters should be chosen based on changes in the spectrum morphology.
\item The set should represent the RVS data treated by the pipeline. Because in \gdr we treat only \grvs $\leq $12 and because for bright stars, the morphology of the spectra is important, we used two sequences, one with \FeH= 0  and \Teff\ = 3100, 3500, 4000, 4500, 5000, 5500, 6000, 6500, 7000, 7500, 8000, 9000, 10\,000, 15\,000, 20\,000, 25\,000, 30\,000 and 35\,000 K and the other with  \FeH=$-1.5$ and \Teff  = 3100, 3500, 4000, 4500, 5000, 5500, 6000, 6500, 7000, to represent cool stars with low metallicity.
\item For the gravity we used Table 3.3 from \cite{2012A&A...543A.100R}, suggesting that most stars are giants with $\log g \sim 3$ (but slightly depending on \Teff). 
The main-sequence value of $\log g$ = 4 is used for the hotter
stars.
\item A rotational velocity = 0.0 ${\rm km\,s}^{-1}$ was assumed when converting these synthetic spectra into templates.
\end{itemize}

The atmospheric parameters of the 28 spectra that were selected from the auxiliary synthetic spectral library are also tabulated in Sect. \ref{sssec:detAP}. 

The 28 synthetic spectra were transformed into templates to be used in {\tt DetermineAP}. The transformation of the synthetic spectra into templates was made as described in Sect.~\ref{sssec:generateTemplate}, with the difference that here the spectra were convolved with a Gaussian LSF with $R$ = 11~500, independent of the CCD and FoV. This simple template simulation is sufficient for the {\tt DetermineAP} purposes, which is to compute the maximum value of the correlation peak for a template selection and not to derive a radial velocity.

\subsection{Auxiliary \grvs}\label{sssec:auxgrvs}
The \refgrvs~magnitudes of the stars contained in this file were used as the reference to estimate the \intgrvs\ zeropoint (Sect. \ref{sssec:grvszp}). We computed \refgrvs~for 113 686 \textit{Hipparcos} stars
using the formula  \refgrvs~$ = V - 0.0501 - 1.1667*(V-I) + 0.0052*(V-I)^{2} + 0.0011*(V-I)^{3}$, where $V$ and $I$ are the magnitudes in the {\it Hipparcos} catalogue  \citep{2010A&A...523A..48J}. The magnitude estimation using the $V$ and $I$ bands is more precise than the one that we have in IGSL, and for this reason, \refgrvs\ (and not \extgrvs) was used for the zeropoint estimation. 


\section{\bf Calibration of the RVS instrument \label{sec:calibration}}

The pipeline for this first data release provides the first-order calibrations necessary to treat bright star spectra that do not overlap with spectra in other windows. More calibrations will be necessary in future versions to treat fainter stars and blended spectra.

\begin{figure}[h]
\begin{center}
\includegraphics[width=0.5\textwidth]{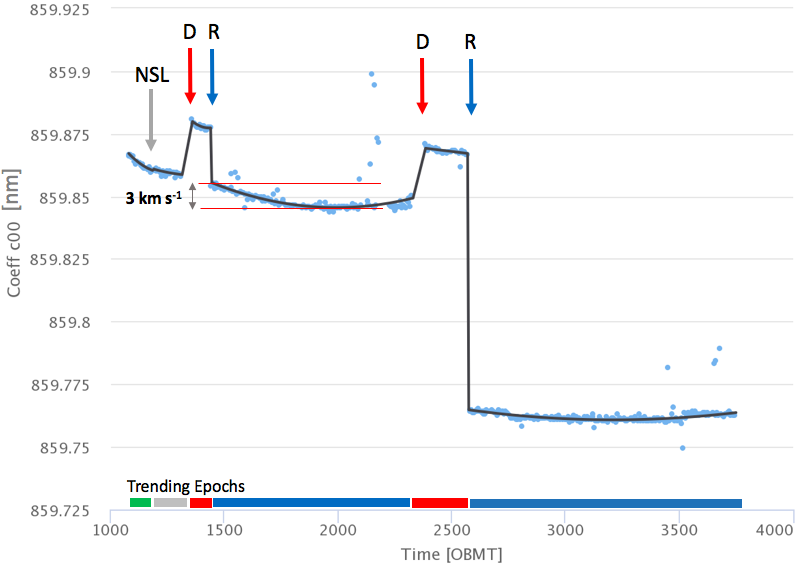}

\caption{Wavelength calibration model: Trending function of the coefficient $C_{00}$, representing the wavelength-calibration zeropoint, shown for the CCD in row 5, strip 16, and FoV1 (black line). The timescale on the abscissa is expressed in OBMT (revolutions, Sect.~\ref{sec:inputdata}).
The blue points are the values of $C_{00}$ obtained at each calibration unit. The arrows indicate breakpoints. The most important discontinuities are due to optics decontamination (red arrows, at OBMT 1317 and 2330.6) and to the re-focusing events (blue arrows, at 1443.9 and 2574.6). An additional breakpoint was set at the transition between EPSL and NSL, at OBMT 1192.13 (grey arrow). The \gdr data set that covers the time between OBMT 1078 (25 July 2014) and 3750.5 (23 May 2016) was divided into six trending epochs that are indicated with colour bars along the abscissa:  shortly after each decontamination, a re-focusing was needed, and the trending epochs between these two events (the red bars) are short. The two red horizontal lines delimit the variation of the wavelength zeropoint over one trending epoch. These variations are typically small, $< 3$ \kms ($\sim 0.3$ pixels).}
\label{fig:C00trend}
\end{center}
\end{figure}

The most important calibration necessary for estimating \RV\ is the wavelength calibration. It is included in the pipeline, which calculates the associated calibration parameters and their spatial and temporal dependence. The payload module decontamination and refocusing operations produce discontinuities in the basic angles that are reflected in the wavelength calibration coefficients. Figure \ref{fig:C00trend} shows the temporal evolution of the coefficient $C_{00}$ representing the wavelength calibration zeropoint. 
An additional smaller discontinuity is produced by the change in the \gaia\ scanning law from the ecliptic pole scanning law (EPSL) to the nominal scanning law (NSL). \gaia\ was operated in EPSL during the first weeks of operations (between 25 July to 21 August 2014), with the spin axis in the ecliptic plane and the FoV scan through both the south and north ecliptic poles \citep{DR1-DPACP-18}. Based on these discontinuities, five breakpoints were defined that correspond to these events, and the entire calibration data set was separated into six \textup{trending
epochs}. A calibration model was produced for each trending epoch.

In addition to the wavelength calibration, the bias non-uniformity (bias NU) effect (Sect. \ref{sssec:biasnu}), which produces jumps in the spectra (see Fig. \ref{fig:winextcalspectrum}) that compromise the \RV\ estimation even in bright stars, needed to be calibrated. 
The calibration coefficients were computed by a dedicated RVS
pipeline, and were shown to be stable for the \gdr purposes \citep{DR2-DPACP-29}, so no temporal variation was implemented.

The other calibrations that were necessary to treat the spectra but that had only second-order impact on the \RV\ estimation of bright stars were not included in the pipeline. These are the line-spread-function in the AL direction (LSF-AL), the CCD blemishes and saturation level, and straylight. The calibration model used for these was determined off-line{\it  } using a limited data set, or the pre-launch data. We describe here the calibration models we used to obtain the \gdr \RV, both those produced by the pipeline, and those produced off-line.

\subsection{Electronic bias non-uniformity}\label{sssec:biasnu}

\cite{DR2-DPACP-29} (see also Cropper et al. 2018) described the bias NU effect, its calibration procedure, the resulting parameters and their temporal evolution in detail. A constant voltage offset is applied to the CCDs prior to digitalisation. This electronic bias is needed to avoid negative signal values and wrap around zero-digitised units (ADU) at low signal levels. This constant bias level is measured from the prescan data that are acquired every 70 minutes in bursts of 1024 samples. The constant (uniform) bias calibration consists of extracting the median signal from the bursts of prescan data for each CCD and of applying a linear interpolation to these data to obtain the bias offset that is to be subtracted from the spectra at any time. 

In addition to this constant bias, owing to the manner in which the \gaia\ CCDs are operated, the electronic bias applied to an acquired sample prior to digitisation contains an unstable component such that the bias value applied to a given sample depends on the details of the read-out process, in particular on the number of {\it \textup{flushes}} (fast skipping of the inter-window pixels before the sample), the position of the {\it \textup{glitches}} caused by interruptions of the serial clocks during the operation of the parallel clocks, and the common baseline (the difference between the prescan bias and the image area bias). These features are known collectively as the bias NU anomaly and are described in detail by \cite{DR2-DPACP-29}.

The calibration of the RVS bias NU is done by a dedicated pipeline using well-defined sequences of samples that are called special sequences and are acquired during special calibration periods about three to four times per year, using CCD gates to hold back photoelectrons.  The result of the calibration is a set of coefficients
that model the flushes and glitches for each CCD. These coefficients are used to correct the RVS spectra (Sect. \ref{sec:extraction}).

The special sequences were acquired twice during the 22 months covered by the \gdr data, and the RVS bias NU effect appeared to be stable \citep{DR2-DPACP-29}. For this reason, only one set of coefficients (the most frequently tested one that was
obtained during commissioning) was used to correct the entire data set, and any temporal variation of the bias NU anomaly was neglected. 


\subsection{Scattered light calibration }\label{sssec:scatterlight}

\begin{figure}
\begin{center}
\includegraphics[width=0.5\textwidth]{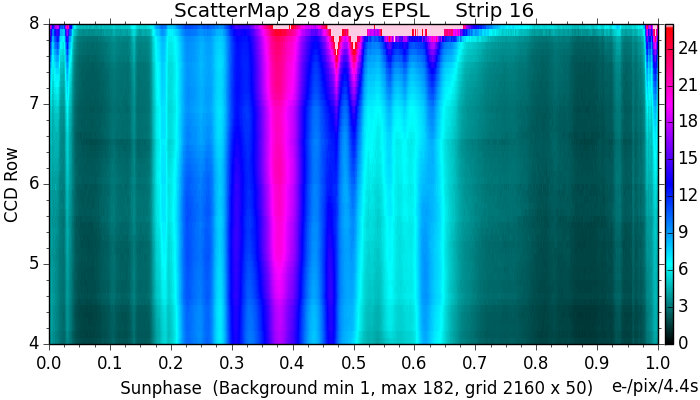}
\caption{Background map (scatter map) used in the pipeline for the four CCD rows in strip 16 (strip 16 in the \gaia\ focal plane is numbered 2 in the RVS focal plane). On the y-axis is the across scan (AC) direction plotted as CCD row number, while the x-axis is the \gaia\ solar phase angle. This scatter map is obtained using the VOs acquired during the 28 days of operations, when \gaia\ was in EPSL mode (first trending epoch). The scatter map is dominated by the solar straylight. The units are \epix (not divided by the 4.4 second exposure time).
\label{fig:scattermap}}
\end{center}
\end{figure}

After launch, it was discovered that the readout windows of the RVS spectra were contaminated with a diffuse background dominated by solar straylight \citep{DR2-DPACP-46}. 
The straylight contamination varies over time and also with the position on the focal plane, but a large part of it follows a quite stable pattern based on the satellite rotation phase. The typical pattern is shown in Fig. \ref{fig:scattermap}.

The background level was measured using the {\it \textup{virtual objects}} (VOs), which are windows containing only background signal (no sources) that are acquired every 2 s and cover the CCDs uniformly in the AC direction. 
The background level measurements obtained from the VOs are used to build a 2D map (Fig. \ref{fig:scattermap}) dominated by the solar straylight, which is called the scatter map. The scatter map is used to obtain the background level to subtract from the RVS spectra, depending on their AC position on the CCD and on
the rotation phase of the satellite during their acquisition.

The scatter map used in the pipeline was produced off-line using the VOs acquired only during the first 28 days of \gaia\ operation, when \gaia\ was in EPSL mode, in the first trending epoch. It is used to correct the entire data set, and temporal variations
are not considered. 

To produce the scatter map, the software proceeds in two steps:

\begin{enumerate}
\item{\tt MakeScatter} is responsible for treating the single VOs. First, the VO spectra with truncated windows, 2D windows, or windows that are affected by a CCD defect are discarded to remove other signal contamination from the background measurements. Then, the VO spectra are corrected for the bias NU. 
The background flux level is calculated as the median flux of all valid samples in the VO spectrum after excluding the samples that are affected by cosmic rays or saturation. These results are stored in data files (one per VO), called {\it Scatter}, that also contain all the information needed to create the scatter map: the CCD and the AC position of the VO, including the CCD row as well as the solar phase of \gaia\ at the acquisition time. 

\item {\tt MakeScatterMap} uses all the accumulated Scatter data and combines them based on their position, to create one 2D map. A map covers one revolution time in $x$ and all CCD AC positions in $y$. It is composed of a grid with 2160x50 cells (10 seconds AL and 157 AC pixel per grid cell), chosen in order to have a sufficient number of background Scatter measurements per cell (18 measurements). For each Scatter measurement, the solar phase angle is calculated using the satellite attitude. The AC cell index is calculated as $ y  = \frac{{rowfrac} - 4}{4.} * ngridcells $, where {\it rowfrac} is the AC location, subtracting 14 prescan pixels and half a pixel to refer to pixel centre, then divided by 1966, which is the number of exposed CCD pixels. For each grid cell, the straylight level is then the median value of all corresponding Scatter background measurements, and the associated uncertainty is the standard error of the median.

\end{enumerate}

 The correction applied to the spectra is a simple subtraction of a constant level from the entire spectrum and does not significantly affect the \RV\ estimation for these bright stars. However, it affects the magnitude \intgrvs~ (Sect. \ref{sssec:intmag}). The background level calibration was not a priority for this pipeline version and will be improved for DR3.

\begin{figure}
\begin{center}
\includegraphics[width=0.5\textwidth]{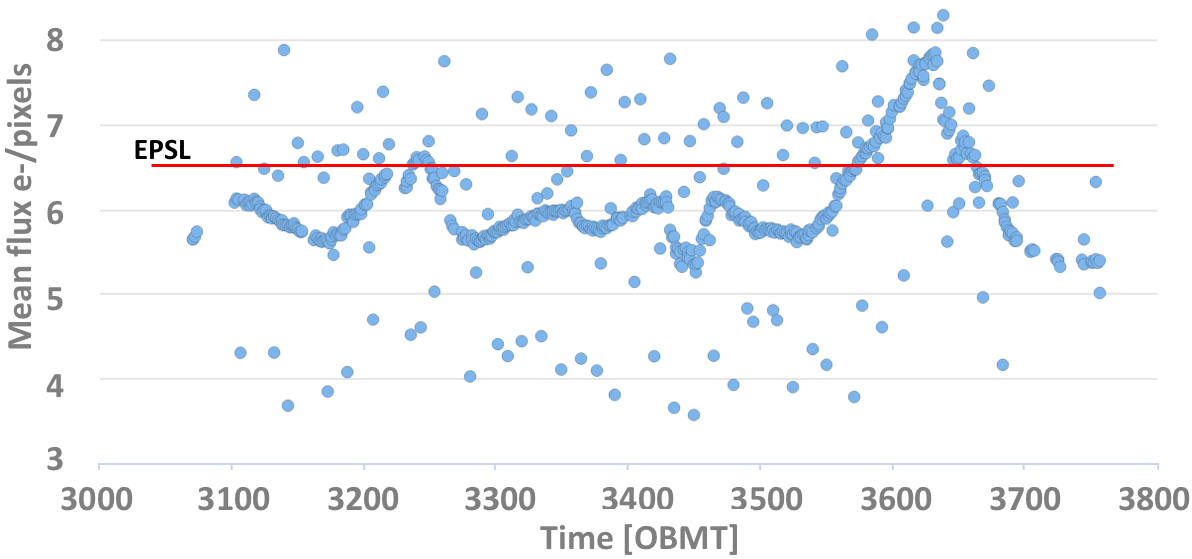}
\caption{Temporal evolution of the mean level of the scatter map in CCD row 6 strip 16 during the last period covered by the \gdr data set. The timescale is expressed in OBMT, and the time interval shown corresponds to 12 December 2015 and 23 May 2016. Each blue point represents the mean background level over 30 hours. It is computed using the scatter map obtained with the VOs acquired in 30 hours (outliers are from VO incompleteness). The units are e$^-$/pix, not divided by the 4.4 s exposure time. The red horizontal line at 6.54 \epix\ is the mean level of the scatter map obtained with the EPSL VO data and is used to process the whole DR2 data. The peak at OBMT 3600 corresponds to a scanning of the Galactic plane by \gaia\ and arises from the contribution to the straylight from the high density of external (i.e. non-solar) sources in these scans.
\label{fig:scattertrend}}
\end{center}
\end{figure}

In order to quantify the consequences of neglecting time variations and using the constant scatter map obtained off-line with the EPSL data to process all of the DR2 data, the average background level of the EPSL map was compared with the average level of the background over the last five months of data covered by \gdr. During this period, the daily pipeline (which started operation in December 2015, with the primary goal of tracking and verifying the status of the RVS) estimated and monitored the average background level every 30 hours. Figure \ref{fig:scattertrend} shows the variation in average background level (in CCD strip 16) between December 2015 and May 2016 compared with the average EPSL scatter map level from August 2014. The figure shows that the mean background level during EPSL was higher than during the period between December 2015 and May 2016. This may be explained by the fact that EPSL receives a constant contribution from the galactic plane and/or zodiacal light that is due to the stable orientation of the \gaia\ spin axis (not doing any precession).
 
\subsection{Wavelength calibration}\label{sssec:wavecalib}

The RVS is not equipped with wavelength-calibration lamps as these are difficult to implement in TDI operation. Hence a self-calibration approach is adopted, where some of the RVS observations themselves are used to calibrate the instrument
\citep{2007sf2a.conf..485G}. The basic principle of the RVS wavelength calibration is that the wavelength value associated to a sample can be expressed as a function of the FoV coordinates $(\eta, \zeta)$ of the source at the time when the sample crosses the CCD fiducial line (Sect.~\ref{sssec:ingestion}).

Each FoV and CCD of the RVS is calibrated independently and continuously using the data acquired during time intervals of 30 hours in order to obtain one wavelength calibration every 30 hours. This time period, over which the calibration stars and the characteristics of the instrument are assumed to remain constant, defines the wavelength calibration unit (CaU) duration. 

The time evolution of the wavelength calibration (zeropoint, dispersion, and ariations in the spatial direction) is then modelled in the trending module, using the information obtained in all CaUs. 

\subsubsection{Selection of calibrator stars}

\begin{figure}[h!]
\begin{center}
\includegraphics[width=0.5\textwidth]{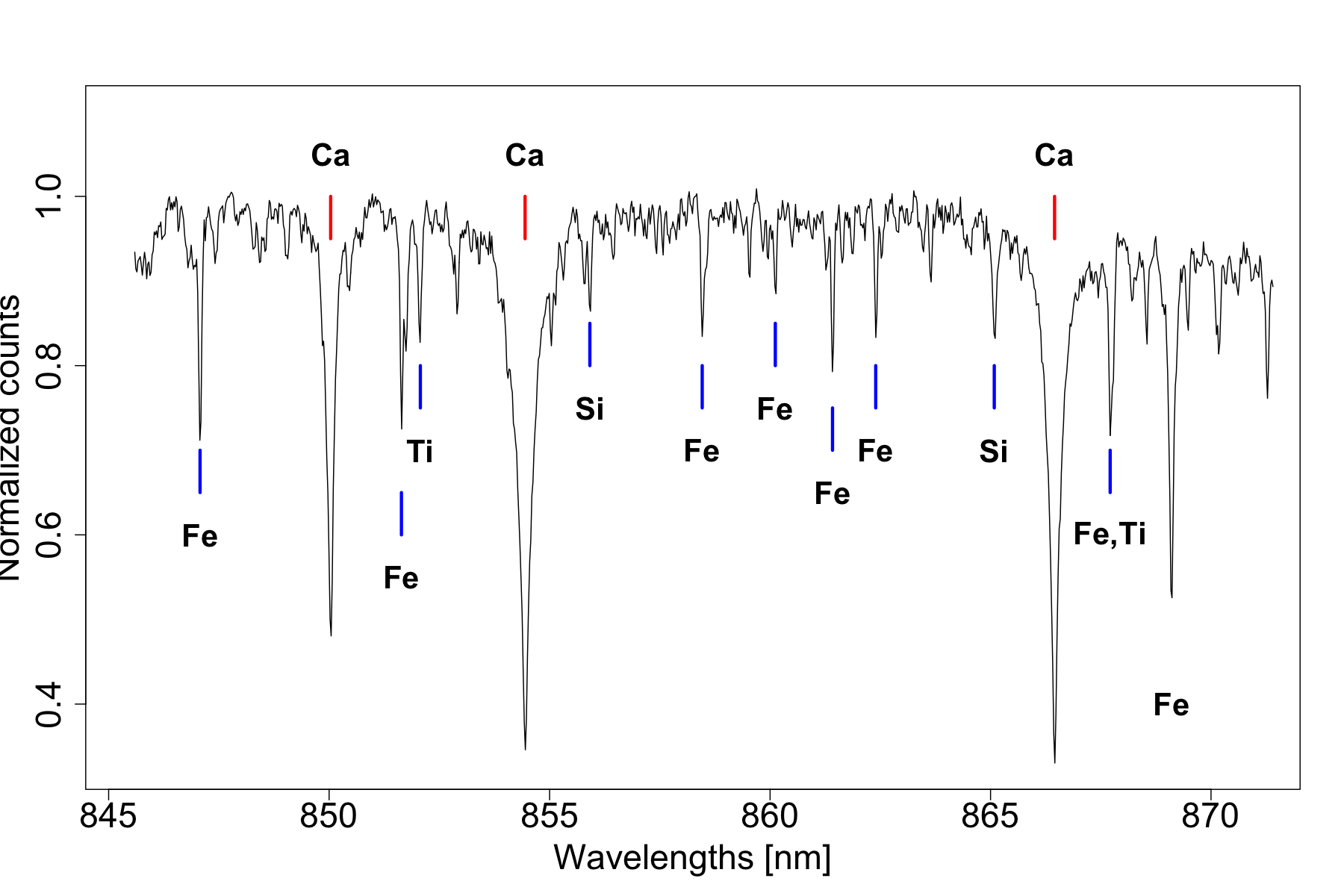}
\caption{RVS spectrum of a bright star used for wavelength calibration. The calcium triplet lines are indicated in red, and some Fe and Si lines are indicated in blue.  
\label{fig:starwithlines}}
\end{center}
\end{figure}

The first step of the wavelength calibration process is to select
those star spectra from the RVS observations acquired during the CaU duration that present several well-detected narrow lines to be reference line candidates. In practice, we selected the spectra of stars with \Teff\ in the interval $4500  < T_{\mathrm{eff}} <  6500$ K, and \grvs\ $\le 9$ (Fig. \ref{fig:starwithlines} shows a typical spectrum of a calibrator star). The calibration stars (those that satisfy the selection conditions) also include \RV\ standard stars with well-known, stable radial velocities from the auxiliary radial velocity catalogue (Sect.~\ref{sssec:auxstd}).
These are used to fix the wavelength calibration zeropoint. The spectra have passed through the {\tt Calibration Preparation} workflow (Fig.~\ref{fig:workflows}) and were cleaned (Sect.~\ref{sec:extraction}) and calibrated with the first-guess model. The wavelength range of each input spectrum was reduced to avoid including leading/trailing edge data that might confuse the algorithm. In addition, the suitable template spectrum was selected from the auxiliary synthetic spectral library (Sect.~\ref{sssec:synthspe}). The identified template lines were used to `predict' where the corresponding RVS spectrum lines might lie. 

\subsubsection{Identification of the reference lines}

The second step is to identify the reference lines suitable for wavelength calibration in these
calibration spectra and to associate the triplet ($\lambda_{\mathrm{rest}},\eta,\zeta$) with each of them. There is no definitive list of reference lines. Rather, the list of lines is determined by the template assigned to each calibrating spectrum. Blended lines and lines presenting cosmic rays or saturation are discarded. These are the tasks of a software module called {\tt Centroiding}, which works in the following way:

\begin{itemize}
\item Detection of the deepest lines in the spectra. The 20 deepest lines of the spectrum are detected (they are sorted by decreasing depth), and estimates of the location of their centroids are evaluated. These first centroids are expressed in integer sample values, with the centre of the bluest sample of the spectrum associated with the value 0. The Ca lines are easily identified;
they are normally the deepest lines in the spectra, but the code takes into account that sometimes the \ion{Fe}{i} line, which
lies close to the red wing, is observed as a deeper line than some
of the triplets. 

\item Identification of the reference lines based on their predicted positions. The prediction algorithm uses the Ca line sample positions identified during the line detection phase and the wavelengths of these lines in their template counterpart to compute a wavelength scale.
Next, the algorithm takes each template line in turn and identifies a sample within the RVS spectrum that represents the notional centre of the matching line ($s_0$). It then assigns four samples from either side of this centre to construct a nine-sample wide line. If the centre sample is the deepest flux sample of the nine, then the line is deemed a reference line; otherwise, it is discarded. 
\item Deriving rest wavelength. 
The rest wavelength is derived via cross-correlation between the detected reference lines and their template counterparts. 
A local calibration relation is defined:
\begin{equation}
\lambda = k \times s + \delta\lambda  =  k \times s' + \lambda_0
\label{eqn:lambda_rest}
,\end{equation}
where $s$ is the sample, $\lambda$ is the wavelength corresponding to the sample $s$, $k$ is the sampling factor which takes into account the spectral resolution element relative to the sample size, $s'$ is the changed frame of the observed line ($s'=s-s_0,$ where $s_0$ is the discrete centroid of the line), $\delta\lambda$ is the wavelength shift of the local calibration law, and $\lambda_0$ is the wavelength location of the discrete centroid ($\lambda_0 = k \times s_0 + \delta\lambda$).  The best match between the template and the observed lines is found by a classical cross-correlation, that is, by computing the optimal parameter $\lambda_0$ associated with the sample $s_0$.  Because the template is at rest, $\lambda_0 = \lambda_{\textrm{rest}}$. The rest wavelength is derived as follows:

\begin{itemize}
\item Shift the local calibration relation in wavelength by modifying the parameter $\lambda_0$ of the local calibration relation.
\item Calibrate the observed line using the local calibration relation.
\item Re-sample the at-rest template to the sampling of the observed line. 
\item Compute and analyse the correlation coefficient using the RVDir method (see Sect. \ref{sssec:cu6spe_RVDir}).
\item Using an interpolation algorithm, the centroid locations $(s_0, \lambda_{\textrm{rest}})$ are expressed in field angles ($\eta_s,\zeta_s$). 
\end{itemize}
\end{itemize}
At the end of this process, each reference line of each spectrum has a position and a rest wavelength: ($\lambda_{\mathrm{rest}},\eta_s,\zeta_s$).

\subsubsection{Determination of the dispersion function}

The third step of the wavelength calibration process is to determine the spectral dispersion for the CaU in question, using the reference line triplets ($\lambda_{\mathrm{rest}},\eta_s,\zeta_s$) produced in the previous step. The dispersion function is represented as a second-order bivariate polynomial,

\begin{equation}
\lambda_{\mathrm{observed}} = \lambda_{\mathrm{rest}}\left(1+\frac{V}{c}\right) = \sum^2_{m=0}\sum^1_{n=0}C_{mn}\eta_{s}'^m\zeta_{s}'^n
\label{eqn:lambda_sfull}
,\end{equation}

where
\begin{itemize} 
\item  $\lambda_{\mathrm{observed}}$ is the wavelength associated with the centre of the sample;
\item $C_{mn}$ are the unknown calibration coefficients;
\item $V$ is the unknown {\it Gaia}-centric radial velocity of the calibration star;
\item $\eta'_s$ and $\zeta'_s$ are the field angle FoV coordinates of the calibration stars at the fiducial time of the CCD sample, shifted and scaled. Shifting and scaling are needed to improve the stability of the solution;
\item $\eta'_s$ is shifted and scaled $\eta_s$: $\eta'_s = s_{\eta}(\eta_s - \eta_{0})$;
\item $s_{\eta}$ is the $\eta$ scale factor, which is the same for all configurations: $s_{\eta} = 1000$;
\item $\zeta'_s$ is shifted and scaled $\zeta_s$: $\zeta'_s= s_{\zeta}(\zeta_s - \zeta_{0})$;
\item $s_{\zeta}$ is the $\zeta$ scale factor, which is the same for all configurations: $s_{\zeta} = 10$;
\item $\eta_0$ and $\zeta_0$ are the pivot points, they are fixed for each FoV and CCD and correspond to the projection of the middle of the readout register into the $\eta, \zeta$ plane for a wavelength of 860.5 nm.
\end{itemize}

To improve the stability of the solution, we set to zero the coefficients $C_{21}$, $C_{02}, C_{21}, \text{and } C_{12}$ in equation \ref{eqn:lambda_sfull}, and obtain

\begin{equation}
\lambda_{\mathrm{observed}} = \lambda_{\mathrm{rest}}\left(1+\frac{V}{c}\right) = C_{00} + C_{10} \eta'_s + C_{20} \eta'^{2 }_s+ C_{01} \zeta'_s + C_{11} \eta' _s\zeta'_s
\label{eqn:lambda_s}
.\end{equation}

To avoid the degeneracy because a shift in the dispersion law can be compensated for by a shift of the radial velocities of the calibrator stars, the auxiliary ground-based standard stars with known radial velocity (Sect.~\ref{sssec:auxstd}) are used to fix the zeropoint. The standard stars are also part of the calibrator stars because they satisfy the conditions required to be selected as calibrator. $V^{\mathrm{ref}}$ is their radial velocity, transformed into {\it Gaia}-centric radial velocity.

The calibration coefficients and the radial velocities ($V^s$) of the calibration stars are derived simultaneously by a least-squares
fit, minimising the following function $\chi_k^2$: 

\begin{equation}
\resizebox{0.95\hsize}{!}{$
\chi_k^2  =  \sum^{N_{\mathrm{spectra}}(k)}_{r=1}\sum^{N_{\mathrm{line}}}_{l=1}\left[{\lambda_{r,l} + \left(\frac{\lambda_{r,l} V^{\mathrm{ref}}}{c}\right)} + {\left(\frac{\lambda_{r,l} V^{r \neq\mathrm{ref}}}{c}\right) - \sum^2_{m=0}\sum^1_{n=0}C(k)_{mn}\eta_{r,l}'^m\zeta_{r,l}'^n}\right]
$}
\end{equation}

where

\begin{itemize}

\item $N_{\mathrm{spectra}}(k)$ is the number of spectra observed during the calibration unit $k$;

\item $N_{\mathrm{line}}$ is the number of reference lines in each spectrum;

\item $C(k)_{mn}$ are the calibration coefficients for the calibration unit $k$ ($C_{02}, C_{21},\text{and } C_{12}$ were set to zero);

\item $\lambda_{r,l}$ is the measured rest wavelength of the reference line from {\tt Centroiding} (\textup{dependent variable});

\item $V^{\mathrm{ref}}$ is the known radial velocity of a ground-based standard; it is shifted to \textit{Gaia}-centric by subtracting the \gaia\ barycentric velocity correction (dependent variable);

\item $V^r$ is the unknown \textit{Gaia}-centric radial velocity of the star in the $r^{\rm{th}}$ spectrum; 

\item $\eta_{r,l}'$ and $\zeta_{r,l}'$ are the FoV angular coordinates of the
calibrator star (corresponding to the $r^{\rm{th}}$ spectrum) at the time associated
with the location of the centroid of the $l^{\rm{th}}$ reference lines in
the $r^{\rm{th}}$ spectrum (independent variable).

\end{itemize}

The minimisation of the $\chi_k^2$, equivalent of resolving a linear system is accomplished using the {\tt Single Value Decomposition} algorithm, implemented in the {\tt Efficient Java Matrix Library} (\url{https://ejml.org/wiki/index.php?title=Manual}).

\subsubsection{Modelling temporal variations}
In the last step, the temporal variations of the wavelength calibration are modelled.
The wavelength calibration process is repeated for all the CaUs of the trending epoch. After the coefficients for the whole epoch are obtained, a trending module analyses the long-term variation of each calibration coefficient, and performs a curve fit that best describes the emerging trends. The type of function that best fits the temporal trend of $C_{00}$ and $C_{10}$ is a second-order polynomial, while a simple median value is used to model the trend of the other coefficients. 

The trending functions are the final results of the wavelength calibration and are used to calibrate all the spectra  acquired at any time and at any position. Fig. \ref{fig:C00trend} shows, as an example, the calibration values and the trending function (black line) of the coefficient $C_{00}$ that represents the wavelength-calibration zeropoint for one of the CCDs of the leading FoV, FoV1, during the time covered by the \gdr data.

\subsubsection{Automated verification}
The quality of the dispersion functions obtained for each CaU and for each FoV and CCD is verified by the {\tt Automated Verification} module. 
First, the number of the standard stars and of 
the calibrator stars used to compute the solution is verified (we need to have a sufficient number of standard stars covering 
the $\zeta$ coordinate in the CCD). Then, the residuals between the wavelength of the reference lines obtained by the 
line centroid algorithm and those obtained by applying the dispersion law are computed:
$\Delta_{\lambda} = \lambda_{\rm{calibration}} -\lambda_{\rm{rest}}(1+\frac{V^s}{c}),$
where\begin{itemize}
\item $\lambda_{\rm{calibration}}$ is calculated using the dispersion law (Eq.~\ref{eqn:lambda_s})
and the field angles $\eta_s$ and $\zeta_s$ of the reference lines;
\item $\lambda_{\rm{rest}}$ is the rest wavelength of the reference line;
\item $c$ is the velocity of light in vacuum;
\item $V^s$ is the velocity of the star, obtained with the wavelength calibration.
\end{itemize}

The temporal variations of the calibration coefficients and the trending functions are inspected by eye. The final quality estimation of the wavelength-calibration model is provided in Sect. \ref{sssec:avsta}, after the extraction of the radial velocities of the standard stars, by comparing the results obtained with the RVS data with the auxiliary ground-based standard values.


\subsection{Line spread function (LSF-AL)}\label{sssec:lsfAL}

The \gaia\ point spread function (PSF) is approximated by the cross product of the AL and the AC line spread functions (LSF). The LSF-AL and LSF-AC calibrations are not implemented in this pipeline version. The LSF-AC calibration will be implemented for DR3 and will be used for deblending the spectra and to estimate the flux loss out of the windows. The LSF-AL, on the other hand, is needed in this pipeline version. It contains information on the resolution of the RVS spectra and is used to convolve the synthetic spectra to generate the templates. The LSF-AL calibration was therefore  estimated off-line, using the 28 days of EPSL data, and was used for the data acquired before the first decontamination at OBMT 1317. For the remaining data, the LSF-AL calibration was derived using on-ground data.  In future data releases, the LSF calibration will be included in the pipeline and the estimation will be improved.

The RVS resolution element is monitored daily in each configuration in 
the first-look (FL) pipeline \citep{DR1-DPACP-7} since the beginning of the \gaia\ operations in order to detect defocusing. In the FL, the resolution element is determined by measuring the width of some unresolved Fe lines via correlation with a mask. The FL diagnostics show that the resolution element is in general very close to the nominal value (3 pixels, corresponding to R=11\,500), with a maximum degradation of $\sim 3$\% owing, in general, to gradual de-focus. The degradation of the resolution element appeared much more important, $\sim 20$\%, at the beginning of the operation before OBMT 1317, when the \gaia\ optics suffered from high contamination by water ice. The high contamination is also shown by bright values of the magnitude zeropoint in Fig. \ref{fig:zptrend}.

\begin{figure}
\begin{center}
    \includegraphics[width=0.5\textwidth]{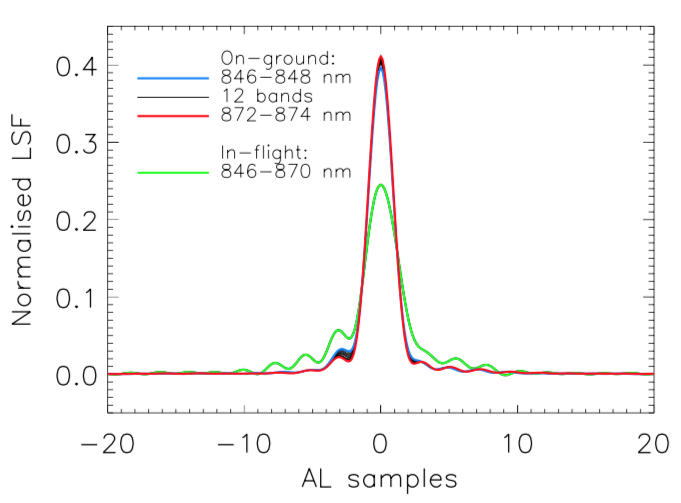}
    \caption{LSF-AL model used in the pipeline for the CCD in row 6 strip 16 and FoV1. The {\it \textup{in-flight}} LSF (green line) is computed using RVS data obtained before the decontamination at OBMT~1317 and shows a significant degradation compared to the nominal, {\it \textup{on-ground}} LSF. The resolution degradation was $\sim 20$\% and was recovered after the decontamination. The {\it \textup{on-ground}} LSF is estimated in 15 wavelength bands. The shortest and longest wavelength bands are coloured blue and red, respectively; the two curves almost overlap, showing the weak dependence of the RVS LSF-AL on wavelength, which is neglected in the \textup{in-flight }model.}
    \label{fig:lsfal}
    \end{center}
  \end{figure}

Two LSF-AL models were estimated, both using principal component analysis \citep{LL:LL-084}: one is the on-ground LSF-AL, which
was used to generate the templates of all the RVS spectra acquired after the decontamination at OBMT 1317;  the other is the in-flight LSF-AL,  which was calculated using the EPSL data (before the 1317 decontamination). This was also used to generate the template for the RVS spectra acquired before OBMT 1317 to take into account the strong contamination of the optics. No other time variation was considered. The two LSF models are shown in Fig.~\ref{fig:lsfal}. 

The on-ground LSF-AL model profiles that were used after OBMT~1317
were produced for each CCD and FoV. They were based on 15 wavebands, each of 2~nm, covering the range 846-874~nm (there are 15 LSFs per CCD and per FoV in total). Time variations and inter-CCD AC variations were neglected.

The in-flight LSF, before the 1317 decontamination, was derived for each CCD and FoV. It was  based on one waveband centred at 858 nm and covering 24 nm (there is one LSF per CCD and per FoV
in total).  It is independent of wavelength, and time variations and inter-CCD variations were neglected. It was modelled using the PCA as a linear combination of eight basis functions:
\begin{equation}
\label{eq:lsfpca}
LS\!F =  \sum^7_{n=0} h_{n} H_{n}
,\end{equation}
where $h_{n}$ are the linear combination coefficients of the basis functions $H_{n}$. The basis functions were selected from a set of on-ground basis functions \citep{LL:LL-084} in order to model the nominal LSF profiles adequately (which were chosen to represent all the data configurations). 

Deriving the observed LSF is equivalent to finding the coefficients $h_n$. 
To do this, we performed a least-squares fit of the linear combination of the basis functions convolved with a high-resolution ground-based spectrum of the observed star. 
Ground-based spectra of about 1200 bright, constant stars were collected for this purpose. They were extracted from the NARVAL and the ESPADON archives\footnote{NARVAL and ESPADON are both high-resolution spectro-polarimeters, with $R$ =  65\,000 - 85\,000. NARVAL is installed on the 2m Telescope Bernard Lyot at Pic du Midi and ESPADON at the 3.6m CFHT.} to represent the RVS observations of these stars with high fidelity. For the purpose of the LSF-AL calibration, the auxiliary synthetic spectra cannot be used, as they do not provide a sufficiently good match with the real spectra. 

The observed RVS spectra ($O$) were modelled as a convolution
of the LSF with the relevant ground-based high-resolution spectrum
with high
signal-to-noise ratio (S/N)  $S$ shifted to the radial velocity
of the RVS observations
and resampled to the RVS observation,
\begin{equation}
\label{eq:convolve}
    p_\lambda O = S*LS\!F(h_0,h_1, ..) 
,\end{equation}

where:  $p_\lambda$ is a wavelength-dependent scaling factor to allow for  flux and also slope and curvature differences;   
$LS\!F(h_0, h_1, \ldots)$  is the LSF function, and $\text{the
asterisk}$  is the convolution operator. 

Using Eq. (\ref{eq:lsfpca}), we can rewrite Eq. (\ref{eq:convolve}) as

\begin{equation}
\label{eq:lsq}
p_\lambda O = \sum^7_{n=0} h_n X_n
,\end{equation}

with

\begin{equation}
\label{eq:xn}
X_n = S*H_n
.\end{equation}

As Eq. (\ref{eq:lsq}) is of the form of a general linear least-squares fit, we can solve for the linear combination coefficients $h_n$ with $X_n$ as the basis functions of the fit. $p_\lambda$ is a second-degree polynomial of the form
$p_\lambda = a_0 + a_1 \lambda + a_2  \lambda ^2$, 
and minimises the differences between the auxiliary spectrum and the
RVS observation. The $a_n$ are obtained by linear least-squares solving for each
sample $i$ in the RVS spectrum the set of equations $(S*H_0)_i/O_i = a_0 + a_1  \lambda_i + a_2  \lambda_i ^2$.

\subsection{\grvs~zeropoint}\label{sssec:grvszp}

The pipeline provides the calibration and temporal evolution of the \intgrvs\ zeropoint ($ZP_{G_{\rm{RVS}}}$), which is useful for monitoring the contamination during the period covered by the \gdr data set. The $ZP_{G_{\rm{RVS}}}$ is estimated for each CCD and FoV and for each CaU (30 hours). The calibration is made using the {\it Hipparcos} stars listed in the auxiliary file as reference (Sect. \ref{sssec:auxgrvs}), for which we have estimated the reference magnitudes \refgrvs.  
\begin{itemize}
\item First, the RVS spectra of the calibration stars (Sect. \ref{sssec:auxgrvs}) were selected; they were already cleaned and calibrated in wavelength. Spectra with a window size AC < 10 were discarded to limit the flux loss, which typically left about 2000 stars per CaU.
\item The $ZP$ is estimated for each spectrum by $ZP^{\rm{spec}} = G_\mathrm{RVS}^{\mathrm{ref}}  +  2.5 \log(TotFlux)$,
 where \refgrvs\ is the magnitude coming from the auxiliary file, and $TotFlux$ is the integrated flux between 846 and 870 nm, divided by the exposure time for one CCD (4.4 s).
 \item The zeropoint, $ZP_{G_{\rm{RVS}}}$, for each CCD and FoV is the median of the values $ZP^{\rm{spec}}$, obtained with the spectra observed in the relevant configuration. 
 The associated uncertainty is given by their robust dispersion: 
 $\sigma_{ZP} = \frac{P(ZP^{\rm{spec}}, 84.15) - P(ZP^{\rm{spec}}, 15.85)}{2}$, 
 where $P(ZP^{\rm{spec}}, 84.15)$ and $P(ZP^{\rm{spec}}, 15.85)$ are the 84.15$^{th}$ and the 15.85$^{th}$ percentiles of the distribution of $ZP^{\rm{spec}}$.
\end{itemize}

\begin{figure}
\begin{center}
\includegraphics[width=0.5\textwidth]{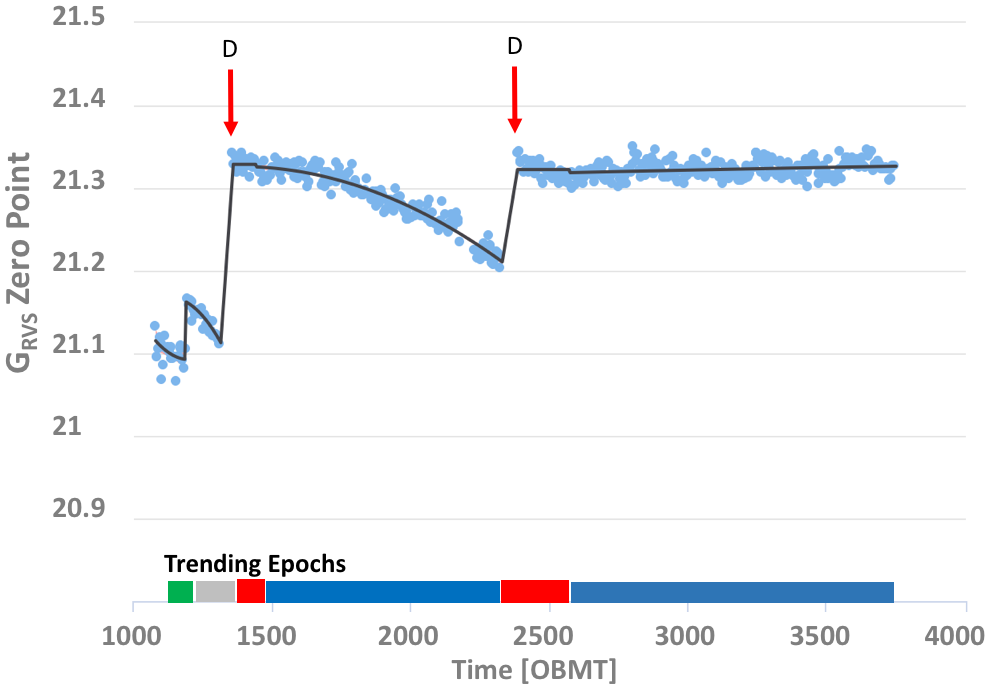}
\caption{\intgrvs\ zeropoint temporal evolution (for the CCD in row 6, strip15 and Fov1). The red arrows indicate the decontamination events that resulted in an improvement of the \intgrvs\ zeropoint. After the decontamination at OBMT 2330.6, there is no degradation of the RVS transmission, and the decontamination events become rarer over mission time. The trending epochs are indicated in
the same way as in Fig. \ref{fig:C00trend}. During the first trending epoch, a wrong catalogue was used, and the zeropoint data obtained in this epoch should be offset by $\sim+0.08$ magnitudes.
\label{fig:zptrend}}
\end{center}
\end{figure}

The temporal evolution of the $ZP_{G_{\rm{RVS}}}$ was modelled with second-degree polynomial trending functions (as was done for the wavelength calibration), see Fig. \ref{fig:zptrend},
and the value of the $ZP_{G_{\rm{RVS}}}$ can be obtained at any transit time. It is used to estimate
the magnitude \intgrvs (Sect. \ref{sssec:intmag}). 

\section{\bf Cleaning and reducing the RVS spectra\label{sec:extraction}}

This section describes the process responsible for transforming the RVS windows into cleaned and calibrated spectra. This process is applied to the spectra to be used for calibration (\textup{i.e.
}in the workflow {\tt Calibration Preparation} of Fig.~\ref{fig:workflows}) using initial, first-guess calibrations, and is then applied to all the spectra, this time using the final calibrations (this is done in the {\tt FullExtraction} workflow of Fig.~\ref{fig:workflows} in charge of preparing the spectra to be used to extract the radial velocities). 

The spectra entering this process have already undergone the on-board processing \citep{DR2-DPACP-46}, and the IDT window reconstruction \citep{DR1-DPACP-7} (Sect. ~\ref{sec:inputdata}). They have also passed the {\tt Ingestion} step (Sect.~\ref{sssec:ingestion}), where all the relevant information for the downstream processing is extracted from the input data. 

\begin{figure}[h!]
\begin{center}
\includegraphics[width=0.5\textwidth]{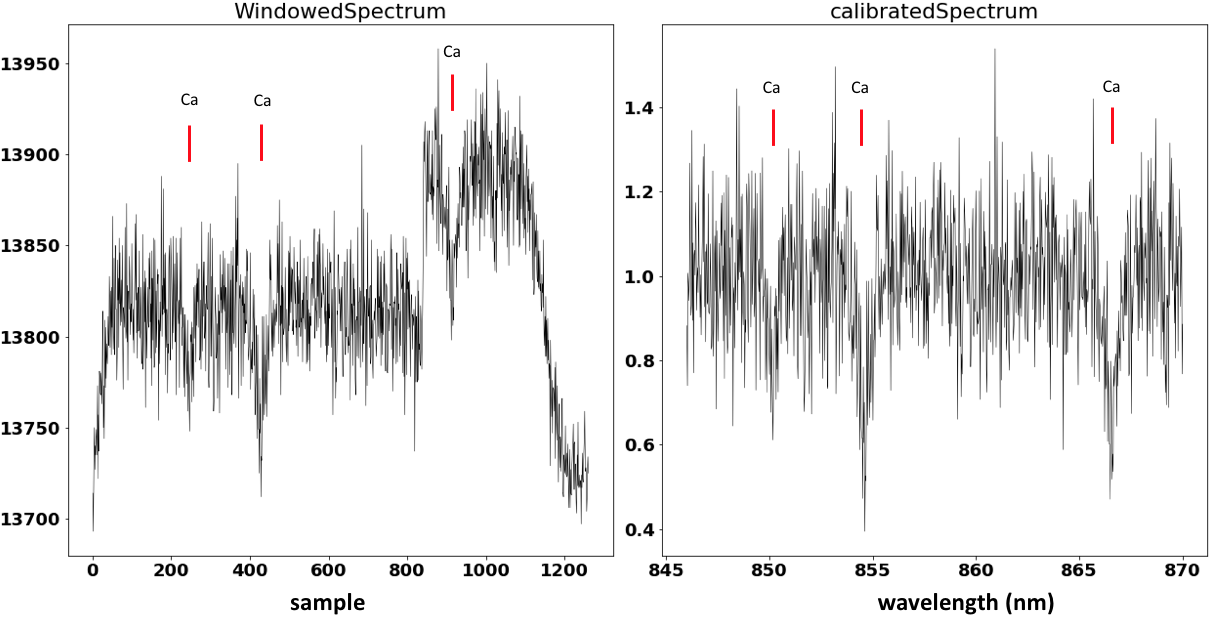}
\caption{Different reduction stages of a spectrum from a 4.4 s exposure on a single CCD. (Left): Raw spectrum. Y is in ADU (analogue-to-digital units), X is in samples (one sample is 1 AL x 10 AC pixels), the biasNU effect is visible as an offset after sample 820. The \ion{Ca}{ii} triplet lines are indicated by red lines. (Right): Spectrum calibrated in wavelength, normalised to the pseudo-continuum, and samples on the leading/trailing edges are discarded. The wavelength range of the calibrated spectra is [846-870] nm. The atmospheric parameters that have been determined by {\tt DetermineAP} (Sect.~\ref{sssec:detAP}) and associated with the star are \Teff\ = 6000 K, \logg\ = 3.5, and \FeH = 0. The internal magnitude is \intgrvs=10.2 (Sect.~\ref{sssec:intmag}), and the $S/N \sim 9$.}
\label{fig:winextcalspectrum}
\end{center}
\end{figure}

Figure \ref{fig:winextcalspectrum} shows a spectrum at two processing stages: at input, before bias NU correction, and at output, cleaned,
and calibrated in wavelength, with the edges cut and normalised. 

\subsection{Raw spectra cleaning}\label{sssec:cleaning}

\subsubsection{Electronic bias correction}\label{sssec:biascorr}

The first step is to correct the flux in the raw spectrum, still in ADU, for the uniform bias
derived from the prescan data and for the non-uniform bias offset (Sect.~\ref{sssec:biasnu}). 
The bias NU correction procedure involves the reconstruction of the readout timing for each sample of the CCD, see \citet{DR2-DPACP-29}. In Fig. \ref{fig:winextcalspectrum} we show a spectrum that is
affected by bias NU.

During the bias correction procedure, the saturated samples (ADU = 65\,535) are identified and flagged. The spectra containing saturated samples are then excluded from the processing, which
removes {\bf $\sim 0.3\%$} of the spectra.

\subsubsection{Gain and dark current correction}

After the bias removal, the spectral flux is transformed into photoelectrons using on-ground gain values. The nominal (on-ground) dark current level of 
2.80 $\times$ 10$^{-4}$ electrons/pixel/s is then subtracted.

\subsubsection{Background removal}

The background level is subtracted from each spectrum using the ScatterMap obtained as described in Sect.~\ref{sssec:scatterlight}. Based on the AC location of the RVS window, and the \gaia\ Solar phase angle, the position of the window in the ScatterMap is found and the corresponding straylight level is subtracted from the spectrum. 
The spectra falling in ScatterMap regions where the straylight level is higher than 100 $e^-pix^{-1}$ in the 4.4 s exposure are flagged and excluded from the processing. In total, another $\sim 0.3\%$ of spectra have been excluded for this reason.

The straylight level is assumed here not to vary with time. The consequence of neglecting time variations, and using, for the entire data set, the ScatterMap obtained using the data obtained only during EPSL, is that, in general, the level of straylight subtracted from the spectra is too high (see Fig. \ref{fig:scattertrend}), and many spectra, at the faint magnitude end had a negative total flux after the straylight subtraction. They are excluded from the processing. $\sim 9\%$ of spectra are excluded because of negative total flux.

Note that only the offset is subtracted, but the noise in the signal, induced by the straylight, can not be subtracted and has the effect of reducing the effective S/N of the spectra. The S/N degradation is more important for the faint stars. As an example, considering only poisson and readout noises, the S/N per sample in one CCD spectrum of a \grvs\ = 12 star, would be of $\sim 4.5$ without straylight and it is degraded by a factor of 2, when the straylight level is 7 \epix\ and by a factor of 3 when the straylight level is of 22 \epix\ (these are typical values of the ScatterMap in Fig.~\ref{fig:scattermap}).

\subsubsection{Cosmetic Defects}

The list of pre-launch cosmetic defects is taken from a list provided by the CCD manufacturers, e2v\footnote{e2v is called Teledyne e2v as of March 2017.}. No attempt has been made to detect new defects in this pipeline version, nor to measure the column response non uniformity (CRNU). The RVS spectra with pre-launch CCD defects are flagged and excluded from the processing, and {\bf$\sim 1\%$} of the spectra have been excluded. Except for the defects, the CCD column response is considered to be uniform. This approximation is justified for the purposes of this pipeline version by the pre-launch measurements showing that the CCD CRNU is typically less than 1\%.

\subsubsection{Spectra extraction and cosmic ray removal}

The 2D window spectra are extracted and corrected for cosmic rays using the Horne optimal extraction algorithm \citep{1986PASP...98..609H}. The 1D spectra are extracted on board \citep{DR2-DPACP-46}. The cosmic ray correction algorithm compares the three CCD spectra of each transit to identify and correct the cosmic ray hits. The spectra are calibrated in wavelength (Sect.~\ref{sssec:applywavecal}) and normalised to their pseudo continuum (Sect.~\ref{sssec:contnorm}), in order to be directly comparable.
The wavelength samples of the 3 spectra are compared to detect the outliers above 5 $\sigma$ from the median level  and flagged as cosmic ray hits. The flux in the cosmic ray samples is substituted with interpolated values of the nearest samples from the spectra of the other two CCDs.

\subsection{Spectra wavelength calibration}\label{sssec:applywavecal}

This algorithm computes the wavelength scale that is to be applied to each spectrum.
The wavelength calibration coefficients  $C_{mn}$ are computed at any time and position, using the trending functions (Sect.~\ref{sssec:wavecalib}, Fig.~\ref{fig:C00trend}). Each spectrum sample has associated its FoV $(\eta,\zeta)$ coordinates and uncertainty (Sect. \ref{sssec:ingestion}). The wavelength $\lambda_s$ of the sample $s$ is

\begin{equation}
\lambda_s = C_{00} + C_{10}\eta' + C_{20}\eta'^2  + C_{01}\zeta' + C_{11}\eta'\zeta' 
\label{eqn:lambda_s}
,\end{equation}

where $\eta'$ and $\zeta'$ are the shifted and scaled field angles as in Eq.~(\ref{eqn:lambda_sfull}).

The uncertainties on $\lambda_s$ resulting from the propagation of the uncertainties on the coefficients and on the field angles are found to be underestimated. We therefore estimated the wavelength calibration uncertainty {\it \textup{a posteriori}} based on the median dispersion of the single-transit \RVtr\ results for the bright constant stars in the auxiliary data. This uncertainty is estimated to $\sim 400$ \ms (see Sect. \ref{sssec:avsta}), corresponding to $\sim 0.0011$ nm. It is propagated to the single-transit \RVtr\ uncertainty, but is not used to compute the uncertainty on the all-transits combined \RV, published in \gdr.
Instead, the uncertainty is estimated using 
the dispersion among the individual single-transit \RVtr.

After applying the wavelength calibration, the wavelength range of each spectrum is reduced to avoid including the wings of the band-pass filter, which might disturb the \RV\ cross-correlation algorithms. All samples falling outside the wavelength range (846-870 nm) are discarded. Figure~\ref{fig:winextcalspectrum} (right) shows an example of a calibrated spectrum.

\subsection{Internal magnitude estimation}\label{sssec:intmag}
The magnitude \intgrvs is estimated for each CCD spectrum by
\begin{equation}
G_{\rm{RVS}}^{\rm{int}} = -2.5 \log (TotFlux) + ZP_{G_{\rm{RVS}}}
\label{eq:grvsZP}
,\end{equation}
where
\begin{itemize}
\item $TotFlux$ is the total flux from the star measured in the spectrum (cleaned as described in Sect. \ref{sssec:cleaning}), between 846 and 870 nm,  divided by the CCD exposure time.
\item $ZP_{G_{\rm{RVS}}}$ is the zeropoint (Sect. \ref{sssec:grvszp}) computed at the transit time in the same CCD and FoV of the observed spectrum. The zeropoint at the transit time is computed using the trending function. 
\end{itemize}
The estimation of $TotFlux$ on the RVS spectra is limited by the missing appropriate background level estimation (Sect. \ref{sssec:scatterlight}), which needs to be subtracted to estimate the flux level of the source, and by the LSF-AC calibration, which is necessary to estimate the source flux loss outside the readout window. As a result, the magnitude \intgrvs\  that is produced has an insufficient quality for publication in DR2.  Nevertheless, it is used to exclude stars from DR2 whose flux is too low (i.e. with \intgrvs $\geq$ 14).

\subsection{Continuum normalisation}\label{sssec:contnorm}

The spectrum is normalised to its pseudo-continuum, which is determined using a polynomial fitting. First the spectrum is fitted with a polynomial of degree two. The stellar lines are iteratively rejected by sigma clipping, that is,{\it } the pseudo-continuum is set by the pixels within the given interval $[-3, +10]~\sigma$ from the polynomial. 
The polynomial fit and sigma clipping are iterated until convergence on the status of the pixels, as to whether they belong to the pseudo-continuum or not. If convergence is reached, the spectrum is divided by the polynomial.

Spectra with a strong gradient in their pseudo-continuum (e.g. cool stars with a molecular band) are difficult to normalise in this way.
If a positive gradient is detected in the spectrum (cool stars typically exhibit positive gradients) and the windows of the spectrum are not truncated (because when the window is truncated, there is some flux loss outside the window that can result in a gradient), or if the spectrum is too noisy, no pseudo-continuum is computed, and the spectrum is divided by a constant that is the 90$^{th}$ percentile of the fluxes.


\subsection{Atmospheric parameters and template selection}\label{sssec:detAP}

The synthetic spectrum that is associated with each RVS spectrum is selected among the 5256 synthetic spectra of the auxiliary spectral library (Sect. \ref{sssec:synthspe}). The selection is  based on the weighted minimum distance between the atmospheric parameters \Teff, \logg,\ and \FeH that are associated with the star and those associated with the synthetic spectrum, \tempTeff, \templogg\ and \tempFeH. The distance to minimise is:

\begin{equation}
\resizebox{0.95\hsize}{!}{$
D = \frac{|T_{\rm{eff}}-T_{\rm{eff}}^{\rm{tpl}}|}{100} + \frac{|\log g-\log g^{\rm{tpl}}|}{3} + \frac{|{\rm{[Fe/H]}-\rm{[Fe/H]}^{\rm{tpl}}|}}{2}
$}
\end{equation}
where \Teff\ is in K. The weights of 100, 3, and 2 are an ad
hoc{\it } estimate to give more weight to a difference in \Teff\ (on which the morphology of the spectrum depends most) than in \logg\ or \FeH. 

An important task of the pipeline is then to associate appropriate parameters with the stars observed by the RVS. When the atmospheric parameters of the star are known from the literature and are
stored in the auxiliary file (Sect. \ref{sssec:auxparams}), they are associated with the star. When \logg\ or \FeH are not in the auxiliary parameter file, the default values \logg$=4.5$ and \FeH$=0$ are set. 

The majority of the RVS observations, about 85\% in this release, do not have auxiliary associated parameters. 
For these stars, a pipeline module called {\tt DetermineAP} is used to estimate their atmospheric parameters. It works by cross-correlating (Pearson correlation in direct space) the observed RVS spectrum with the 28 templates described in Sect. \ref{sssec:28templates}. The template that gives the highest cross-correlation peak provides the atmospheric parameters to be associated with the 
star\footnote{At fainter magnitudes, \grvs\ $> 12$, which are
not included in this data release, the module does not attempt the cross-correlation, and default parameters are assigned.}. The atmospheric parameters are determined for each transit of the star, and at the end, the parameters that were found for the majority of the transits are associated with the star (Sect.~\ref{sssec:combinetemp}).

To constrain the results of the module {\tt DetermineAP} and for a rough estimate on the uncertainties on the parameter determination and the consequent uncertainties induced on the radial velocity estimations, we computed off-line the atmospheric parameters using {\tt DetermineAP} for a sample of stars for which we also had the parameters from the literature (Sect.~\ref{sssec:auxparams}). A poor estimate of the atmospheric parameters implies that a poor template is associated with the stars and may result in larger systematic shifts in the radial velocity measurements. We then obtained the radial velocity of the stars using the template corresponding to the parameters found by {\tt DetermineAP} and compared them with the radial velocity obtained using the template corresponding to the parameters from the literature.

For this test, the RVS spectra of stars with parameters from the PASTEL compilation \citep{2010A&A...515A.111S}, which are
included in the auxiliary file in Sect. \ref{sssec:auxparams}, were selected from the \gdr data set and processed in {\tt DetermineAP}: 13\,000 stars were observed in $\sim 150\,000$ transits. 

In general, given that \Teff\ is the driving parameter for the presence and strength of spectral lines, the mismatches on \Teff\ between the stars and the templates are expected to produce larger uncertainties on the \RV\ determination than the mismatches on \FeH and on \logg. All the transits were processed with {\tt DetermineAP}. We obtained for each of the 28 templates the residuals $\Delta T_{\rm{eff}} = T_{\rm{eff}}^{\rm{tpl}} - T_{\rm{eff}}^{\rm{cat}}$, between \tempTeff\ from {\tt DetermineAP} and $T_{\rm{eff}}^{\rm{cat}}$ taken from the PASTEL compilation, and the residuals $\Delta V_{\rm R}^{\rm t} = V_{\rm R}^{\rm{tpl}} - V_{\rm R}^{\rm{cat}}$, between the radial velocity obtained using a template with the parameters from {\tt DetermineAP} ($V_{\rm R}^{\rm{tpl}}$), and the radial velocity obtained using a template with the parameters from PASTEL  ($V_{\rm R}^{\rm{cat}}$).  The results of the test are shown in Table \ref{tab:DAPperf}, where
\begin{itemize}
\item $Md(\Delta T_{\rm{eff}})$ is the median of the residuals $\Delta T_{\rm{eff}}$ and indicates the accuracy of the estimation of the effective temperature $T_{\rm{eff}}^{\rm{tpl}}$ of the star obtained by {\tt DetermineAP} (assuming that the PASTEL \Teff\ represents the actual star temperature); 

\item $\sigma(\Delta T_{\rm{eff}})$ is the robust standard deviation of the residuals $\Delta T_{\rm{eff}}$ and indicates the precision of the estimation of the effective temperature of the star by {\tt DetermineAP}. The robust standard deviation is defined as
\begin{equation}
\sigma(\Delta T_{\rm{eff}}) = \frac{P(\Delta T_{\rm{eff}}, 84.15) - P(\Delta T_{\rm{eff}}, 15.85)}{2}
\label{eqn:robuststd}
,\end{equation}
where $P(\Delta T_{\rm{eff}} , 15.85)$ and $P(\Delta T_{\rm{eff}}, 84.15)$ are the $15.85^{th}$ and $84.15^{th}$ percentiles of the distribution of the  residuals $\Delta T_{\rm{eff}}$, respectively.

\item $Md(\Delta$\RVtr) is the median of the residuals $\Delta$\RVtr\ and indicates the shift between the radial velocity obtained with the template found by {\tt DetermineAP} ($V_{\rm R}^{\rm{tpl}}$) and the radial velocity obtained with the template with the PASTEL parameters ($T_{\rm{eff}}^{\rm{cat}}$); it indicates the systematic uncertainty in the estimation of $V_{\rm R}^{\rm{tpl}}$ that
is introduced by the template mismatch.

\item $\sigma(\Delta V_{\rm R}^{\rm t})$ is the robust standard deviation of the residuals $\Delta$\RVtr\, calculated as in Eq.~\ref{eqn:robuststd} and indicates the random uncertainties in the estimation of the $V_{\rm R}^{\rm{tpl}}$ that are introduced by the template mismatch.
\end{itemize}

The radial velocities obtained using the template with the parameters from {\tt DetermineAP} and those obtained with the template with the parameters from the catalogue are in good agreement for $4000 \leq T_{\rm{eff}}^{\rm{tpl}} \leq 6500$ K, despite sometimes large difference in $\Delta T_{\rm{eff}}$. Intermediate-temperature star spectra are dominated by the \ion{Ca}{ii} triplet, and provided the template and the star temperature are both in this range, the template mismatch error in the estimation of $V_{\rm R}^{\rm{tpl}}$ is small. RVS spectra of stars from spectral type B2 to M6 are shown in \citet[][Fig. 17]{DR2-DPACP-46}. 
The {\tt DetermineAP} results degrade for the cool stars where the molecular TiO band appears, and for the hotter stars where in addition to \Teff, \logg\ mismatches play an important role in the morphology of the spectrum because of the appearance of the H lines, which are more pronounced in dwarfs near to the \ion{Ca}{ii} lines. A description of a more detailed study on the template mismatch errors is in preparation by Blomme et al. (in prep.).

In the next data release, the spectroscopic pipeline will include the atmospheric parameters produced with \gaia\ data and more appropriate templates will be assigned to the stars to improve the \RV\ estimation of hot and cool stars.

\begin{table}[h]
\begin{center}
\begin{adjustbox}{width=0.5\textwidth}
\begin{tabular}{rccccc}
\hline\hline
\multicolumn{1}{c}{Template} & \multicolumn{1}{c}{\bf $Md(\Delta T_{\rm{eff}}$)} & \multicolumn{1}{c}{\bf $\sigma(\Delta T_{\rm{eff}})$} & \multicolumn{1}{c}{\bf $Md(\Delta V_{\rm R}^{\rm t}$)} & \multicolumn{1}{c}{\bf $\sigma(\Delta V_{\rm R}^{\rm t})$} &\multicolumn{1}{c}{nb} \\
 ~K~~~dex~~~~dex& K & K & \kms & \kms & \\
\hline\\[-0.3cm]

~3100~~3.0\,~~~~0.0 & $-$1959 & 3870 & 1.23 & 116.46 & 113 \\
~3100~~3.0~~$-$1.5 & $-$1262 &1465 & 1.23 & 2.67 & 61 \\
~3500~~3.0\,~~~~0.0 & $-$1435 & 10896  & $-$0.07  &27.40 & 58 \\ 
~3500~~3.0~~$-$1.5  & $-$1413 & 814 &$-$0.14  & 1.35 & 62\\
~4000~~3.0\,~~~~0.0 & 88 & 500  & -0.05 &0.39 & 1251 \\ 
~4000~~3.0~~$-$1.5 &$-$200 & 426 & $-$0.1 & 0.57 & 261 \\ 
~4500~~3.0\,~~~~0.0 &117 & 324 &0.0  & 0.33 & 4857 \\ 
~4500~~3.0~~$-$1.5 & 113 & 302 & 0.12 & 0.4& 1789 \\ 
~5000~~3.0\,~~~~0.0 & 145 & 222  & $-$0.03& 0.28 & 17829 \\
~5000~~3.0~~$-$1.5 &$-$1050 & 713 &0.05 &0.30 & 5622 \\  
~5500~~3.5\,~~~~0.0 &$-$480 & 314 &$-$0.07  & 0.30& 53491 \\ 
~5500~~3.5~~$-$1.5  & $-$345 & 791 &$-$0.06  &0.49  & 853 \\
~5500~~4.5\,~~~~0.0  & 145  & 302 & $-$0.14 & 0.3 & 6294 \\ 
~6000~~3.5\,~~~~0.0  & $-$508 & 217 & 0.06 & 0.66 & 33297\\
~6000~~3.5~~$-$1.5 & 674 & 726 & $-$0.24  & 0.52 & 1185\\
~6500~~3.5\,~~~~0.0 & $-$280 & 211 & $-$0.12 & 1.47 & 11805\\
~6500~~3.5~~$-$1.5 & $-$425 & 845 &$-$1.0  & 0.92 & 203\\
~7000~~3.0\,~~~~0.0 & $-$3.0    & 591 & $-$3.28  & 11.08  & 3499\\
~7000~~3.0~~$-$1.5 & $-$2348 & 5886 & 4.31 & 8.50 & 234\\
~7500~~3.0\,~~~~0.0 & 200 & 743 & $-$9.02 & 19.40 & 919\\
~8000~~3.0\,~~~~0.0 &$-$1247 & 1099 & 5.57 &  2.58 & 252\\
~9000~~3.0\,~~~~0.0 & $-$1680 & 9708 & 0.14 & 14.66 & 84\\
10000~~3.0\,~~~~0.0 & $-$4380 & 4883 & 0.74 & 2.90 & 243\\
15000~~4.0\,~~~~0.0 & 4714 & 5957 & $-$3.09  & 9.14  & 4403\\
20000~~4.0\,~~~~0.0 & $-$1100 & 4357 & $-$2.24 & 9.22 & 91\\
25000~~4.0\,~~~~0.0 & 2753 & 9574 & $-$8.79 & 48.92  & 77\\
30000~~4.0\,~~~~0.0 & 24340 & 12273 & 3.17 & 285.1  & 17\\
35000~~4.0\,~~~~0.0 & 5000 & 12350 &$-$1.16  & 46.90 & 122\\
 
\hline 
\end{tabular}
\end{adjustbox}
\caption{Performance of the module {\tt DetermineAP} and template mismatch errors. Each of the 28 templates in the first column is identified with its atmospheric parameters (\tempTeff, \templogg\ , and \tempFeH); nb is the number of transits for which the template has been selected.}
\label{tab:DAPperf}
\end{center}
\end{table}

\subsection{Combine template atmospheric parameters}\label{sssec:combinetemp}

{\tt DetermineAP} estimates the atmospheric parameters for each transit of the star. Then, for each star, the parameters found most frequently are selected and the synthetic spectrum with these parameters is used to derive the \RV\ of the star.

These parameters are published in \gdr (Sect. \ref{sec:overview}) and provide information on the synthetic spectrum that was used to obtain the star \RV.  

The transit information is combined in the workflow {\tt Combine Template} (shown in Fig.~\ref{fig:workflows}). Even though it contains only simple functionality, this workflow is technically important because it is the first that uses information from all the transits. It runs after all the data were processed by the other workflows, and together with the following {\tt STAMTA} workflow, works per source. 

\subsection{Template generation}\label{sssec:generateTemplate}

The {\tt GenerateTemplate} module starts from a synthetic spectrum
(Sect. \ref{sssec:synthspe}). The spectrum is convolved with the instrumental profile (Sect. \ref{sssec:lsfAL}),
using the LSF-AL corresponding to the CCD and FoV coordinates of the spectrum.
Then, the spectrum is resampled to a wavelength step of 0.00747 nm, which is ten times
finer than the nominal resolution element of the RVS.
The module can convolve the spectrum with a rotational profile,
but for \gdr, the projected rotational velocity was set to zero.

The wavelength range for the template is [843, 873] nm. This
 is slightly larger than that of the object spectrum in order to ensure than the shifted template in the velocity range
$\pm~1000$ \kms can always be resampled on the object spectrum.
The template spectrum is then normalised (Sect. \ref{sssec:contnorm})
in the same way as the observed spectra.


\section{Deriving the single-transit radial velocity\label{ssec:sta}}

The three CCD spectra corresponding to each FoV transit of a star, with the corresponding synthetic template spectra associated, are analysed by the single-transit analysis ({\tt STA})
set of modules to determine the radial velocity of the star at each FoV transit. This is the radial velocity of the star with respect to the \gaia\ satellite, referred to as the spectroscopic radial velocity, \SRV.

The \SRV \  computed by STA  is then corrected for the \gaia\ barycentric velocity to provide the radial velocity with respect to the barycentre of the solar system, 
\begin{equation}
\label{equ:baryvelocorr}
V_{\mathrm R}^{\mathrm t} = sV_{\mathrm R}^{\mathrm t} + BaryVeloCorr
.\end{equation}
The \RVtr\ obtained at each transit of the source are then combined (Sect. \ref{ssec:mta}) to provide the median barycentric velocity of the source \RV. In \gdr, \RV\ are provided, but not \RVtr.  

\subsection{Single and double stars per transit analysis}
\label{sssec:sta}

\begin{figure}
\centering
\includegraphics[width=5cm]{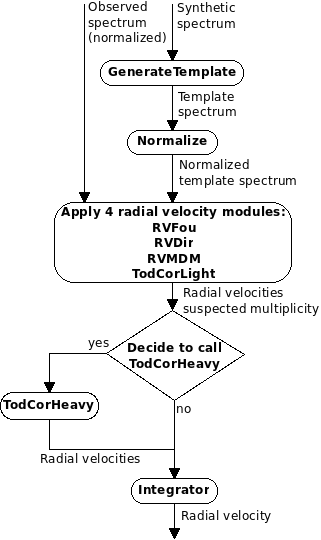}
\caption{Flowchart showing the main steps of the Single Transit Analysis
pipeline. }
\label{fig:cu6spe_STA_flowchart}
\end{figure}

Figure \ref{fig:cu6spe_STA_flowchart} gives a flowchart of the modules that 
make up the {\tt STA} part of the processing. 
Names and acronyms on this figure are explained below.
The main inputs to the {\tt STA} are the observed 
cleaned and calibrated spectra
(a set of three spectra for one transit) 
and the synthetic spectrum (Sect. \ref{sssec:synthspe}) that is chosen to correspond to the 
astrophysical parameters determined in Sect. \ref{sssec:combinetemp}.
From the synthetic spectrum, the templates are generated (Sect. \ref{sssec:generateTemplate})
and normalised (Sect. \ref{sssec:contnorm}) in the same way as the observed spectra.

The normalised observed spectra and template are then processed in
sequence by four or five different radial velocity modules. 
{\tt RvDir} (Sect. \ref{sssec:cu6spe_RVDir}) uses the Pearson correlation 
coefficient, {\tt RvFou} (Sect. \ref{sssec:cu6spe_RVFou}) the standard method
of cross-correlation in Fourier space, and {\tt RVMDM} (Sect. \ref{sssec:cu6spe_RVMDM}) the minimum distance method.
In order to search for binaries, {\tt TodCorLight}  (Sect. \ref{sssec:cu6spe_TodCorLight})
applies a technique equivalent to a 2D cross-correlation, assuming that the
astrophysical parameters of the secondary are the same as those of
the primary.
If {\tt TodCorLight} finds that the spectrum is double-lined, a 
range of astrophysical parameters is explored by {\tt TodCorHeavy} (Sect. \ref{sssec:cu6spe_TodCorHeavy}). For this range, the set of 28 templates listed in Table \ref{tab:DAPperf} is used.
The four or five radial velocity modules each provide their separate results. 
These are then passed through {\tt Integrator} (Sect. \ref{sssec:integrator}), 
which combines them into a single result.

Each radial velocity module was implemented independently, with only a few common
rules. This approach was chosen
to guarantee that the implementation contains the best features
appropriate to it, unhampered by too many constraints on the programming details.

All of the radial velocity modules handle the three CCDs corresponding to one 
transit, and in the process of determining the spectroscopic radial velocity of the star during the transit, \SRV, they also determine 
the spectroscopic radial velocity in each of the three CCDs (\SRVCCD). 
All modules have been designed with the ability to handle the case where information from one or two CCDs is missing. 

\subsection{RVFou}
\label{sssec:cu6spe_RVFou}

{\tt RvFou} implements the standard cross-correlation method for Doppler-shift 
measurement using the Fourier transform \citep[][Sect.~2.3.1]{2014A&A...562A..97D}. 
For cross-correlation, the observed spectrum and the template must be resampled on 
a grid with constant step in log(wavelength) and normalised
to their continuum. Furthermore, ideally, both of their edges should be featureless, 
but this cannot be guaranteed in an automated environment. Therefore, to avoid 
problems with the Gibbs effect on the Fourier transforms,
we apply edge apodisation (5\%) with a cosine bell function after
normalisation. 
This ensures that the spectra behave like continuous periodic functions for the 
Fourier transform.

Fourier transforms are computed using the fast Fourier transform (FFT) 
algorithm found in the {\sc Apache Commons Maths} java library. 
The FFT technique requires that the number of flux bins
in the two data segments (i.e. observed spectrum and template) be a power of
two; within this constraint, the number of samples was optimised and chosen to be a factor of twice
the power of two, which gives the closest (but slightly larger)  number of wavelength bins in the observed spectrum.

The maximum position of the cross-correlation function (CCF) defines the value of the spectroscopic radial velocity for the RVS CCD spectrum in question \SRVCCD. It  is estimated by fitting a parabola to the peak of the
function. Since we cannot refine the wavelength grid at will, we locate the peak by combining the results of 
a three- and four-point parabola fit 
to reduce 
the impact of discretisation, as proposed by \citet{1995A&AS..111..183D}.

Internal uncertainties on the radial velocity measurements are estimated using the 
expression proposed by \citet[][Sect.~2.3]{2003MNRAS.342.1291Z}, that is,
\begin{equation}
\sigma^\mathrm{2}_{sV^\mathrm{CCD}_\mathrm{R}} = 
- \left( {N \,\,\,  {{C^{\prime\prime}} \over {C}} ~~ {{C^\mathrm{2}}   \over {\,\, 1 - C^{\mathrm{2}}}}} \right) ^{-1}
\label{equ:cu6spe_STA_sigcc}
,\end{equation}
where $N$ is the number of wavelength bins, $C$ represents the 
value of the CCF, and $C^{\prime\prime}$ the value of its second derivative, both evaluated at its maximum.

These operations are carried out separately on the three CCD spectra recorded in 
a transit to obtain three \SRVCCD ; for a transit-combined \SRV\ measurement 
and uncertainty estimate, the three 
correlation functions are merged into one, as proposed in 
\citet{2003MNRAS.342.1291Z}. 

\subsection{RVDir}
\label{sssec:cu6spe_RVDir}
In the module {\tt RvDir}, the radial velocity of the input spectrum on a single CCD, \SRVCCD, is 
obtained from the maximum value of the Pearson correlation function (PCF). 
The PCF is defined as

\begin{equation}
C_\mathrm{pc}(v) = \frac{\sum_{n=1}^N (f_n-\bar{f})(t_n(v)-\bar{t}(v))}
{\sqrt{\sum_{n=1}^N (f_n-\bar{f})^2\sum_{n=1}^N (t_n(v)-\bar{t}(v))^2}}  \ \label{equ:cu6spe_STA_PCF}
,\end{equation}
where ${f_n}$ is the flux in wavelength bin $n$ of the observed spectrum, and 
${t_n(v)}$ is the flux of the template spectrum shifted with radial velocity $v$ 
and resampled to the wavelength grid of the observed spectrum. 
Object and template spectrum are both normalised to the continuum. $N$ is the 
number of wavelength bins of the object spectrum; $\bar{f}$ 
and $\bar{t}(v)$ are the averaged fluxes of the object spectrum and of the shifted 
template spectrum, respectively. The radial velocity shift of the input spectrum 
corresponds to the maximum of the PCF. 

For each CCD, the PCF 
is computed in a first step on a coarse velocity grid ranging from $-1000$~\kms\ to $+1000$~\kms\ 
with a step $\Delta v_{\rm coarse} = 10$~\kms and its highest value is identified. 
In a second step, the PCF is computed with a finer velocity grid 
$\Delta v_{\rm fine} = 0.5$~\kms over a reduced range that extends to $\pm~50$~\kms\ 
from the maximum of the coarse array. Thus the PCF for each CCD is obtained as 
an array defined on an irregular grid (steps are either $10$~\kms or $0.5$~\kms).

In order to achieve a sub-step accuracy, a parabola is fitted through the 
highest three points of the `fine' PCF sampling. The maximum of that parabola 
defines the value of \SRVCCD\ for the spectrum at hand. 

An internal uncertainty estimate is obtained using the same expression as for {\tt RvFou}, 
specifically Eq. \ref{equ:cu6spe_STA_sigcc}, where 
$C$ now represents the value of $C_\mathrm{pc}(v)$ and  $C^{\prime\prime}$ its second derivative 
with respect to $v$, both evaluated at the radial velocity \SRVCCD\ corresponding to the maximum value of the PCF. 

To construct the combined PCF, we merge the three single-CCD velocity grids 
into one that contains all of their points. There may be grid points
in the combined grid for which fewer than three values are 
directly available from the previously calculated arrays; such missing values 
are filled in by linear interpolation. Finally, the combined PCF is obtained 
as the mean of the former three PCFs.  The same procedure is applied to the 
combined PCF to obtain \SRV\ and $\sigma_{sV_{\mathrm{R}}^{\mathrm{t}}}$.

\subsection{RVMDM}
\label{sssec:cu6spe_RVMDM}

The {\tt RVMDM} module determines the radial velocity shift that minimises 
the `distance' between the observed spectrum and the template 
that was Doppler-shifted and resampled to the observed wavelength 
grid. The theoretical background for this technique is discussed 
in \citet[][Sect.~2.3.3 and Appendix B]{2014A&A...562A..97D}. For 
convenience, both spectra are normalised. Using the same notation 
as in Sect. \ref{sssec:cu6spe_RVDir}, we define the  $\chi^2$ distance function as
\begin{equation}
C_{\rm md}(v) = \sum_{n=1}^{N} \frac{(f_n- t_n(v))^2}{\sigma_n^2} 
\label{equ:cu6spe_STA_md}
,\end{equation}
where $\sigma_n$ is the uncertainty on the observed flux.

The function is first evaluated on a coarse velocity grid ($-$1000 to 
+1000~\kms, with a step of 20~\kms). A parabola is fitted to the lowest 
three points, its minimum providing the Doppler shift as measured on 
that grid. Then the step is halved; we define a new search range 
centred on the grid point 
nearest to the previous measurement, with twice the 
previous step size as width, and repeat the Doppler-shift measurement.  
This refinement is iterated eight times so that the smallest grid-step size 
is about $\sim 0.08$~\kms, which corresponds to approximately 1\% of a 
wavelength bin. 

An internal uncertainty estimate on this measurement is obtained by 
starting from the best-fit value $C_{\rm md,min}$ and changing the velocity 
until $C_{\rm md}(v) = C_{\rm md,min} + 1.0$, noting
$v_1$ and $v_2$ for the velocities where this occurs. The threshold $1.0$ is 
the $\Delta \chi^2$ value that defines a 68.3\% confidence region for a 
fit with one parameter ($v$). We define the error bar as the maximum of 
$ | v_1 - sV_\mathrm{R}^{{\mathrm t}} | $ and $ | v_2 - sV_\mathrm{R}^{{\mathrm t}} | $ to mimic a 
classical one-$\sigma$ error bar. 

The function $C_{\rm md}(v)$ for the combined CCDs is obtained trivially by 
extending Eq.\ref{equ:cu6spe_STA_md} to a sum over the three CCDs, but this 
requires a common velocity grid whose definition is complicated by the 
occurrence (in the single-CCD grids) of nine different step sizes in regions 
that need not be common, and by the fact that the minimum of the combined 
$C_{\rm md}(v)$ may occur in a region where none of the single-CCD grids was 
refined. Thus the combination may require additional shift and 
resampling operations on part of the data, as well as careful tracking of 
all $v$-values involved. At the end of the calculation, both \SRVCCD\ for each CCD, 
and the transit combined \SRV\ are provided with their uncertainties.


\subsection{TodCor}
\label{sssec:cu6spe_TodCorLight}
\label{sssec:cu6spe_TodCorHeavy}

The three {\tt STA} modules described above
(Sects. \ref{sssec:cu6spe_RVFou}, \ref{sssec:cu6spe_RVDir} and \ref{sssec:cu6spe_RVMDM}) 
are designed to measure the unique Doppler shift exhibited by single-lined spectra. 
However, gravitationally bound systems (binaries or multiple stars)
might generate composite spectra; this is also the case for accidental
confusion on the RVS line of sight. Moreover, two FoVs are observed simultaneously in the same focal plane, which may
give rise to false composite spectra. {\tt TodCor} is dedicated to the detection of 
composite double-lined spectra.

The module was initially designed as an implementation 
of the {\tt TodCor} method first described by \citet{1994ApJ...420..806Z}. 
However, our {\tt STA} algorithm, still called {\tt TodCor}, evolved from this initial 
design to implement the following
functionalities: i) derive the uncertainties on the measured radial velocities and on the brightness ratio, and 
ii) assign a probability that the observed spectrum is double-lined rather than single-lined.
A more detailed description of this new algorithm is in preparation by in Damerdji et al. (in prep.).
Its actions can be described as follows:
\begin{itemize}
        \item Process the input spectrum as a single-star
        spectrum for deriving a single-star model goodness-of-fit.
        \item Process the input spectrum as a double-star
        spectrum for deriving a double-star model goodness-of-fit.
        \item Compare the goodness-of-fit, check the significance
        of the difference (taking into account the different degrees of freedom)
        and decide the nature 
        of the observed spectrum. Thereafter, output the computed velocities 
        ($V_R$ or $V_{R1}$ and $V_{R2}$) of the most likely model.
\end{itemize}

For each transit and for each model, an approximate solution is found for the individual and combined CCD
spectra by assuming equal flux uncertainties.
The input spectra to this method (observed and template) have normalised continua. A minimum $\chi^2$ is derived 
in Fourier space thanks to the Parseval equality 
\citep[][Sect.~12.1]{NumericalRecipes}. 

The second part of the algorithm refines the approximate solution taking into account the flux uncertainties.
It consists of a Levenberg-Marquardt minimisation \citep[][Sect.~15.5.2]{NumericalRecipes}.
For the single-star model, the observed spectrum is assumed to be 
$S(\lambda_i) = P_n(\lambda_i) * T_1(\lambda_i,V_{\mathrm{R1}},\varv_{\mathrm{rot1}})$,
while it becomes 
$S(\lambda_i) = P_n(\lambda_i) * \left[T_1(\lambda_i,V_{\mathrm{R1}},\varv_{\mathrm{rot1}}) + \alpha~T_2(\lambda_i,V_{\mathrm{R2}},\varv_{\mathrm{rot2}})\right]$ in the double-star model, where $P_n(\lambda)$
is a polynomial function with degree $n$, linked to the magnitude of the spectrum ($n \leq 2$). Such a polynomial function is added to the fit to model the effect of tilted spectra in window or stellar reddening. The brightness ratio $\alpha$ is part of
the  parameters of the fit, while the polynomial coefficients are optimised.

The refined solution (single- or double-star model) contains the set of parameters, their uncertainties,
and the goodness-of-fit $\chi^2_D$.
The uncertainties on the parameters are given by the diagonal elements of 
the variance-covariance matrix \citep[][Sect.~15.5.2]{NumericalRecipes}.

The TodCor-like algorithm comes in two flavours: {\tt TodCorLight,} and {\tt TodCorHeavy}.
\begin{itemize}
\item{\bf {\tt TodCorLight}} is responsible for the provisional identification of the composite spectra, which are flagged as
{\it \textup{\textup{suspected multipl}e}}. It assumes that the candidate secondary star has  atmospheric parameters identical to the primary star
(i.e. both templates are the same).\\

\item{\bf {\tt TodCorHeavy}} is triggered 
if {\tt TodCorLight} decides that the star is a {\it \textup{suspected multiple}}.
{\tt TodCorHeavy} loops over a list of atmospheric parameters for both the primary and the secondary stars by 
assuming these parameters to be equal for better numerical stability. This will change in the future version 
of the pipeline, when Todcor will use all the transits together.
{\tt TodCorHeavy} returns the atmospheric parameters of the primary-secondary pair that best fit 
the observed spectrum, together with the derived radial velocities and rotational velocities, 
their uncertainties, and the brightness ratio ($\alpha$). 
\end{itemize}

The decision on the binary nature of the observed star is based  on the comparison of 
$\chi^2_S$ and $\chi^2_D$.  $\frac{\chi^2_S - \chi^2_D}{\chi^2_D}$ is assumed to follow an F-distribution.
The binary model is accepted above a threshold of 0.9 and 0.99865 of the F-distribution CDF for {\tt TodCorLight} and {\tt TodCorHeavy,} respectively. 

For this first version of {\tt TodCor}, some additional detection limits were estimated off-line to minimise false detections. The binary model was not accepted, and the star in this transit was considered as single in the following cases:
 \begin{itemize}
 \item when the separation of the two component radial velocities is $|V_{\mathrm{R1}}-V_{\mathrm{R2}}|<$ 20 \kms 
or  $|V_{\mathrm{R1}}-V_{\mathrm{R2}}| >$ 500 \kms;
 \item when $\alpha < 0.25$  and $|V_{\mathrm{R1}}-V_{\mathrm{R2}}|>$ 40 \kms;
 \item  when $\alpha < 0.35$ and $30 < |V_{\mathrm{R1}} - V_{\mathrm{R2}}| < 40 $ \kms;
 \item when $\alpha < 0.5$ and  20 $< |V_{\mathrm{R1}}-V_{\mathrm{R2}}| < 30$ \kms;
\end{itemize}
where $\alpha$ is still the brightness ratio.

\subsection{Integrator}
\label{sssec:integrator}
The {\tt Integrator} is responsible for combining the results obtained by the various methods into a single result and provides a single spectroscopic radial velocity, \SRV, estimate
for each star transit.
For single stars, there are three different determinations of 
the \SRV\ of the star by the modules {\tt RvDir}, 
{\tt RvFou,} and {\tt RVMDM}. As discussed in \citet[][Sect.~6.3]{2014A&A...562A..97D}, even when in some cases a method provides better results than another, no simple general conclusion can be 
drawn about the algorithms' performance versus astrophysical  parameters. Thus, the simplest approach was chosen, and the final \SRV\ was computed as the median of the \SRV\ derived by modules {\tt RvDir}, {\tt RvFou,} and {\tt RVMDM}. The internal 
uncertainty on the measurement selected as the median was then also the internal uncertainty on the final \SRV, unless there are only two valid radial velocities, in which case their uncertainties were quadratically averaged. 

For binary stars, the situation was simple since only one module, {\tt TodCorHeavy}, determined the radial velocity. The {\tt Integrator} module checked whether this module was launched and had provided two radial velocities (and two uncertainties on them), one per component. When this was the case, {\tt Integrator} stored them in the data model. 

The results for binary stars are not published in DR2, but are used to exclude the binary candidates from the data release.

\subsection{Flagging}
\label{sssec:cu6spe_flagging}

For the \SRV\ , the flag {\it \textup{\textit{isAmbiguous}}}  is set by each of the 
methods (Sects. \ref{sssec:cu6spe_RVFou}, \ref{sssec:cu6spe_RVDir}, 
\ref{sssec:cu6spe_RVMDM}). For this purpose, all radial velocity differences 
between the three CCDs are checked. If these are all smaller than 10 \kms, the 
transit is not ambiguous. Otherwise, if at least one radial velocity 
difference is significantly larger than the uncertainty on that difference, the {\it isAmbiguous} flag is set to true. 
A flag {\it isValid} is set to false by any of the modules if a computational problem is encountered. {\tt Integrator} combines these results and sets {\it isAmbiguous} in the case of single stars when at least two of the three methods have provided ambiguous results, and when two of the three methods have provided an invalid result. It sets the \SRV\ to null when all three methods have provided an invalid result.

\subsection{Automated verification of {\tt STA}}\label{sssec:avsta}

\begin{figure}
\begin{center}
\includegraphics[width=0.4\textwidth]{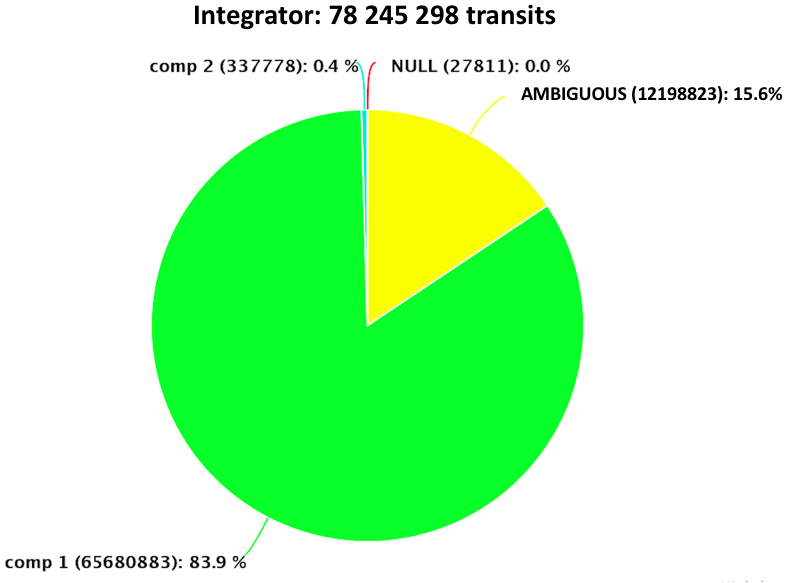}
\includegraphics[width=0.5\textwidth]{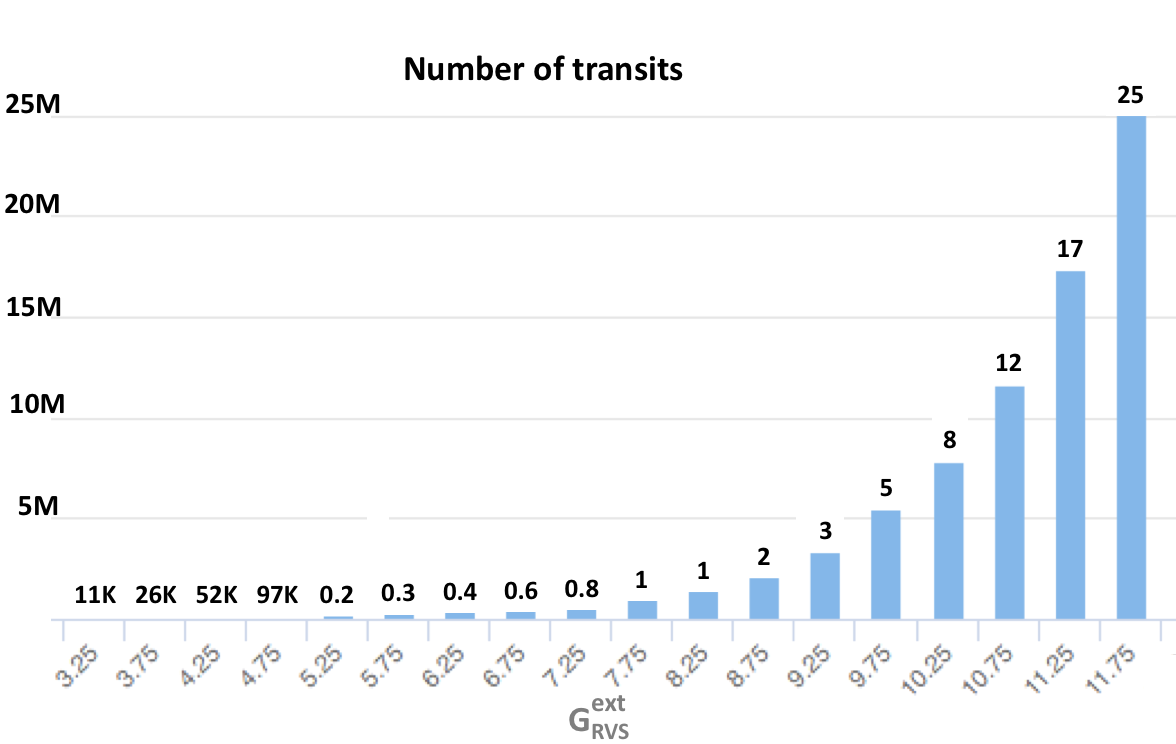}
\caption{{\tt STA} statistics. The total number of transits for which a \SRV\ is obtained is $\sim 78$ million (with  $\sim 28\,000$ null, because invalid in the three {\tt STA} methods {\tt RvFou}, {\tt RvDir,} and {\tt RVMDM}). {\it Top:} Pie chart showing that 16\% of the \SRV\ obtained have been flagged as {\it \textup{ambiguous}}, 84\% of the transits have been detected as single components (comp 1), and in 0.4\% of the transits, multiple-lines have been detected and the transit was flagged as two components (comp 2). {\it Bottom: }Histogram of the distribution of the number of transits as a function of \extgrvs.   
\label{fig:staMultiplicity}}
\end{center}
\end{figure}

Figure \ref{fig:staMultiplicity} shows the number of transits processed by {\tt STA} and their distribution as a function of \extgrvs. {\tt AVSTA} (Fig. \ref{fig:workflows}) verifies the quality of the single-transit radial velocities \RVtr\ (i.e.\textup{} \SRV\ corrected for the barycentric velocity as in Eq. \ref{equ:baryvelocorr}). For this purpose, a verification dataset is extracted from the $\sim 78$ million of \RVtr\ obtained by the pipeline (Fig. \ref{fig:staMultiplicity}), which contains only the stars belonging to the auxiliary radial velocity catalogues (Table \ref{tab:auxradvel}) and covers the magnitude range of the DR2 data (Fig.~\ref{fig:auxradvel-grvs}). These stars are expected to be constant. The radial velocities obtained by the pipeline (\RVtr) are compared with those in the catalogues (\RVref).

\subsubsection{Comparing the results of the {\tt STA} methods}
The first check of {\tt AVSTA} is to ensure that the radial velocities \SRV\ obtained by the three {\tt STA} methods that are combined by {\tt Integrator} (Sect.~\ref{sssec:integrator}) do not present significant systematic differences that might degrade the combined result. The median ($Md$) of the differences between the \SRV\ obtained in the three methods ($\Delta $\SRV = $s$\RV$^{\mathrm{method1}} - s$\RV$^{\mathrm{method2}}$) is computed for the stars in the validation data set. The following results confirm that the systematic differences between the {\tt STA} methods are not significant:  
\begin{description}
\item
[{\tt RvDir}$-${\tt RvFou}]: $Md(\Delta$\SRV) = $-0.02$ \kms; \\$\sigma(\Delta$\SRV) = 0.10 \kms;
\item
[{\tt RvDir}$-${\tt RVMDM}]: $Md(\Delta$\SRV) = $-0.05$ \kms;\\ $\sigma(\Delta$\SRV) = 0.12 \kms;
\item
[{\tt RvFou}$-${\tt RVMDM}]: $Md(\Delta$\SRV) = $-0.04$ \kms; \\$\sigma(\Delta$\SRV) = 0.16 \kms.
\end{description}

\subsubsection{Monitoring the quality of the upstream processing}\label{sssec:avmonitor}
The verification of the radial velocities permits the indirect verification of the performance of the upstream processing, in particular, of the wavelength-calibration performance. For this check, the subset of the verification-dataset containing only the stars belonging to CU6GB-cal (Table \ref{tab:auxradvel}) is used. The \RVref\ of the stars in CU6GB-cal are assumed to be perfectly determined (their uncertainties of $< 0.1$ \kms are neglected), and being used to calibrate the wavelength-calibration zeropoint, should show no systematic shift with the RVS velocities. Any difference is then introduced by the processing, including the upstream AGIS pipeline on which the spectroscopic pipeline is dependent (Sect.~\ref{sec:inputdata}). 

\begin{figure}[h]
\includegraphics[width=0.5\textwidth]{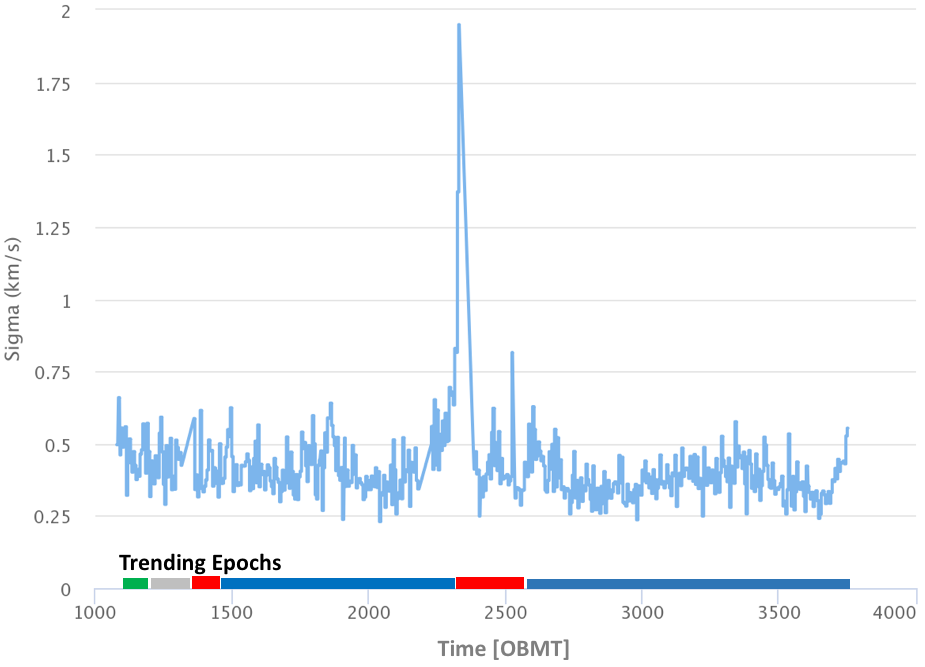}
\caption{Overall precision of the single-transit radial velocity, $\sigma(\Delta$\RVtr), as a function of the observation time (OBMT revolutions) for the standard stars in CU6GB-cal (Table~\ref{tab:auxradvel}). These are bright {\bf FGK-type} stars with \extgrvs $\leq 9$. The trending epochs are indicated on the x-axis. The y-axis shows the robust standard deviation of the residuals, $\sigma(\Delta$\RVtr), calculated over bins of six OBMT. }
\label{fig:STAsigma}
\end{figure}

The precision of the single-transit radial velocities, indicative of random uncertainties, is quantified by the robust standard deviation of the residuals: $\sigma(\Delta$\RVtr) (see Eq.~\ref{eqn:robuststd}).
Figure~\ref{fig:STAsigma} shows the robust standard deviation of the residuals $\sigma(\Delta$\RVtr) calculated over time bins of six revolutions and plotted as a function of time. This diagnostics is used to monitor the precision of the wavelength-calibration over the time covered by the DR2 data set. The peak in the interval OBMT [2324; 2331] was due to poor attitude data from the AGIS solution, resulting in a poor estimate of the field angles of the stars. The \RVtr\ obtained in this interval were not used to produce the combined \RV\ published in DR2. For the remaining time, the precision of the \RVtr\ estimation for the CU6GB-cal stars is $\sim 0.4$ \kms.

\begin{figure}[h]
\includegraphics[width=0.5\textwidth]{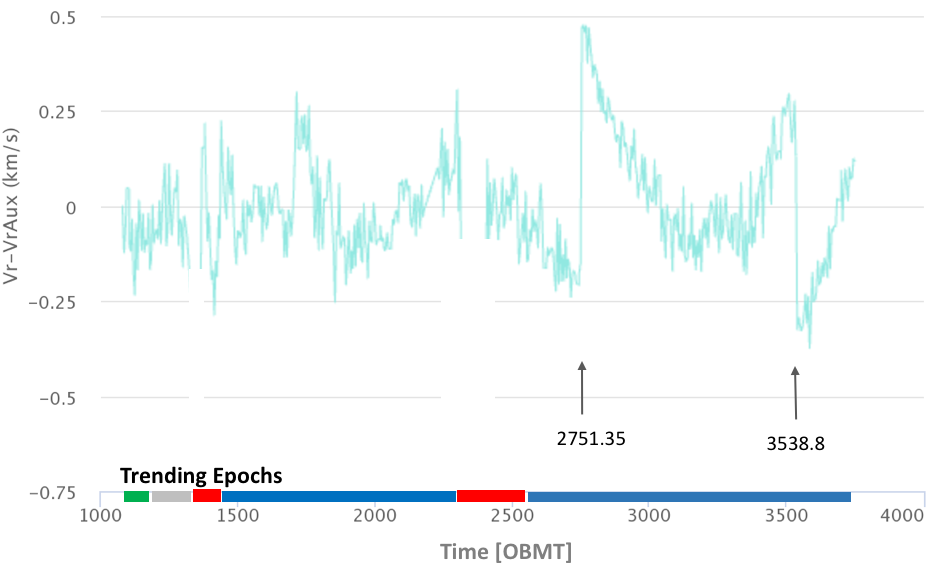}
\caption{Median value of the residuals, $Md(\Delta$\RVtr), as a function of time for the standard stars in CU6GB-cal. This plot is used to monitor the systematic shifts in the wavelength-calibration zeropoint. The trending epochs are indicated on the x-axis. The median is calculated over bins of six OBMT (revolutions). \label{fig:STAresiduals}}
\end{figure}

The accuracy of the single-transit radial velocities, which is
indicative of systematic uncertainties, is quantified by $Md(\Delta$\RVtr), where 
$\Delta$\RVtr~=~\RVtr~$-$~\RVref\  are the differences (residuals) between the \RVtr\ and the \RVref\ in CU6GB-cal. 
Figure~\ref{fig:STAresiduals} shows the median value of the residuals, $Md(\Delta$\RVtr) calculated over time bins of six revolutions and plotted as a function of time to monitor the systematic shifts of the radial velocity zeropoint, which reflect corresponding shifts in the wavelength calibration zeropoint. No explanation has been found for the discontinuities at OBMT 2751.3 and 3538.8. The shifts are smaller than 500 \ms, and being at transit level, do not significantly impact the median \RV\ of the stars affected. Two break-points corresponding to these discontinuities will be set in DR3 to correct for these shifts.

\subsubsection{Dependence on the instrumental configuration}
The subset of the verification-dataset, containing only stars belonging to CU6GB-cal, is used to estimate the systematic differences in the \RVtr\ depending on the CCD row and on the FoV of the transit (the stars are observed in FoV1 and FoV2, the FoVs of the two telescopes, and in general, in a different CCD row at each transit).

The mean, median, and robust standard deviations of the residuals $\Delta$\RVtr\ were obtained over the entire period covered by DR2 for each configuration FoV-CCDrow.
Table \ref{tab:STA-gbsRes} shows the results. The median values $Md(\Delta$\RVtr) indicate the systematic shift of \RVtr\ depending on the configuration of the transit. When no selection was made on the transit configuration (first row of Table \ref{tab:STA-gbsRes}), the overall small systematic shift is $\sim -0.01$ \kms , showing that no zeropoint shift between the RVS \RVtr\ and the \RVref\ in CU6GB-cal is introduced by the processing overall; the temporal evolution of the shift for all configurations together is shown in Fig.~\ref{fig:STAresiduals}.
Systematic shifts are present between the two FoVs (in the opposite sense) and between the different rows. These shifts are small and acceptable for the purposes of this data release. The final radial velocities are the result of the combination between various transits in different configurations, and the shifts are averaged. 
The robust standard deviation ($\sigma(\Delta V_{\mathrm{R}}^{\mathrm{t}})$ in Table \ref{tab:STA-gbsRes}) is indicative of the precision of the measurements \RVtr\ for one transit in a given FoV and row for the stars of spectral type and magnitude typical of CU6GB-cal.

\begin{table}[h]
\begin{center}
\begin{tabular}{lccc}
\hline
\hline\\[-0.3cm]
\multicolumn{1}{c}{FoV-row} &  \multicolumn{1}{c}{$Md$($\Delta$\RVtr)} & \multicolumn{1}{c}{mean($\Delta$\RVtr)} &  \multicolumn{1}{c}{$\sigma(\Delta$\RVtr)} \\
&   \kms & \kms & \kms \\

\hline\\[-0.3cm]
All               & $-$0.01\,~~  & $-$0.04\,~~  & 0.42 \\
\hline\\[-0.3cm]
  FoV1-All    & 0.15 & 0.15 & 0.37 \\
 FoV1-row4 & 0.18 & 0.20 & 0.37 \\
 FoV1-row5 & 0.10 & 0.13 & 0.39 \\
 FoV1-row6 & 0.13 & 0.12 & 0.36 \\
 FoV1-row7 & 0.17 & 0.14 & 0.35 \\
 \hline\\[-0.3cm]
 FoV2-All    &  $-$0.19\,~~ & $-$0.23\,~~ & 0.40 \\
FoV2-row4 &  $-$0.08\,~~ & $-$0.15\,~~ & 0.43 \\  
FoV2-row5 &  $-$0.15\,~~ & $-$0.22\,~~ & 0.45 \\  
FoV2-row6 &  $-$0.20\,~~ & $-$0.17\,~~ & 0.33 \\
FoV2-row7 &  $-$0.31\,~~ & $-$0.41\,~~ & 0.36 \\
  \hline
 \end{tabular}
 \caption{Overall uncertainty depending on the configuration of the observation for the stars in CU6GB-cal. The units are \kms.  Median, mean, and standard deviation of the residuals $\Delta$\RVtr~=\RVtr$-$\RVref (\kms) over the time covered by the DR2. FoV-row are the possible configurations of a star transit (FoV1 and FoV2 are the FoVs of the two telescopes, and row1-4 are the CCD rows, see Fig. \ref{fig:focalplane}).  \label{tab:STA-gbsRes}}
\end{center}
\end{table}

\subsubsection{Comparison with the auxiliary catalogues} 

The entire verification dataset is used (and not only the subset of CU6GB-cal stars) to estimate the performance of the \RVtr\ measurements.
RAVE and APOGEE cover fainter magnitudes than CU6GB-cal (Fig. \ref{fig:auxradvel-grvs}), and permit verifying the accuracy and precision of \RVtr\ up to \extgrvs $\sim 12$. The accuracy and the precision of the \RVtr\ are quantified by $Md$($\Delta$\RVtr) and $\sigma(\Delta V_{R}^t),$ respectively, as described in Sect.~\ref{sssec:avmonitor}. The results for the stars belonging to APOGEE are an overall accuracy of $\sim 0.2$ \kms, an overall precision of $\sim 1.8$ \kms, and for the stars belonging to RAVE, an overall accuracy
of $\sim 0.3$ \kms \ and an overall precision of $\sim 1.4$ \kms.
The uncertainties on the APOGEE and RAVE \RVref\ are larger than
those of CU6GB-cal and contribute to the estimation of the accuracy. These results permitted verifying that overall, the pre-launch requirement on the accuracy on the radial velocities of $\sim 300$ \ms \citep[][their Table 1]{DR2-DPACP-46} was met.


\section{{\bf Combining the single-transit radial velocities}}\label{ssec:mta}

The multi-transit analysis ({\tt MTA}) has the task to combine for each observed star the single-transit radial velocities, \RVtr, to produce the \RV\ published in DR2.
 \RVtr~is combined by applying a median value,
\begin{equation}
V_{\mathrm R} = Md(V_{\mathrm R}^{\mathrm t})
\label{equ:rvmedian}
,\end{equation}
where
\begin{itemize}
\item \RV\ is produced only if the number, $N$, of \RVtr\  (i.e. the number of valid transits) is not smaller than two;
\item the transits for which multiple-component lines have been detected (i.e. those indicated as comp-2 in Fig. \ref{fig:staMultiplicity}) are excluded from the combination;
\item for duplicated transits within 1 s of each other, the transit with the fainter spectrum is excluded. Duplicated transits (i.e. two transits nearby associated with the same source) may occur for bright stars and are the result of false detections due to PSF features.
\end{itemize}

\begin{figure}
\begin{center}
\includegraphics[width=0.5\textwidth]{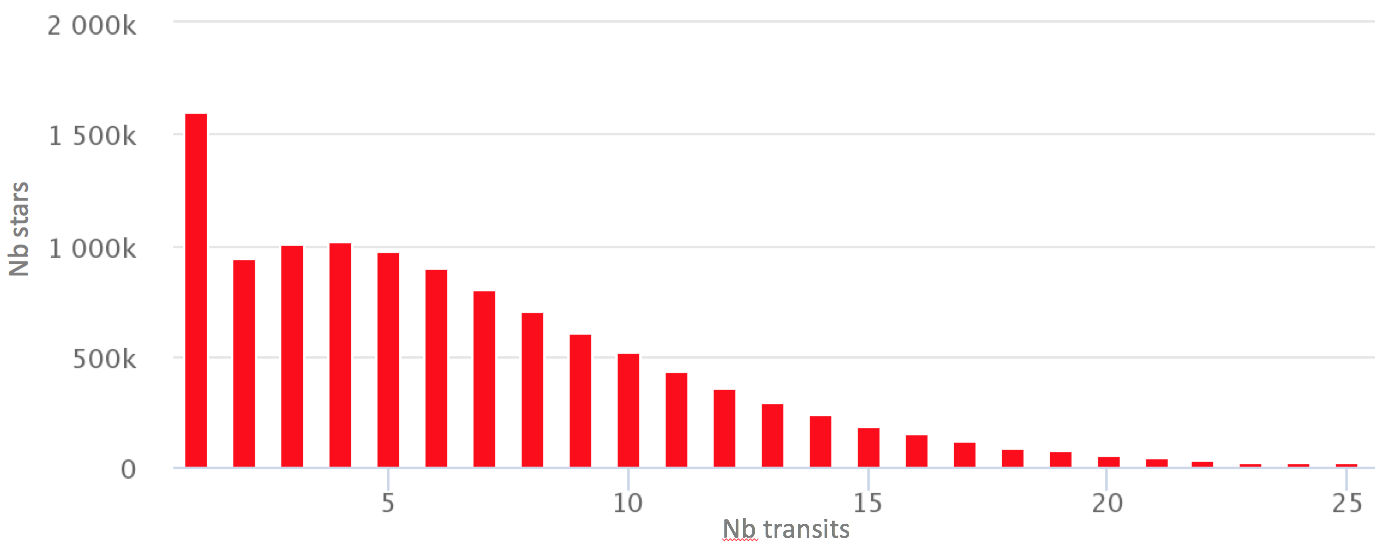}
\caption{Distribution of the number of stars as a function of their number of transits. The x-axis shows the number of transits, and the y-axis shows the number of stars. 1.6 million sources had only one valid transit, and for these sources, \RV\ is not computed. 
\label{fig:mtaTransits}}
\end{center}

\end{figure}

Figure \ref{fig:mtaTransits} shows the distribution of the number of stars as a function of the number of their valid transits.
The uncertainty computed by the pipeline is the standard deviation on the median,\begin{equation}
\sigma_{V_{\mathrm R}}^{\mathrm{MTA}} = \sqrt{\frac{\pi}{2}}\frac{\sigma(V_{\mathrm R}^{\mathrm t})}{\sqrt{N}}
\label{equ:errormedian}
,\end{equation}
where $\sigma(V_{\mathrm R}^{\mathrm t})$ is the standard deviation of the set of \RVtr\ measurements. 

In a post-processing stage, a constant shift of 0.11 \kms \ was
later added to take into account a calibration floor contribution. This shift was estimated by comparing  the internal (Eq. \ref{equ:errormedian})  and
the external precision (Sect. \ref{sec:results}) for a set of constant stars from CU6GB-cal whose radial velocity \RVref\ is known with the best precision. Then, the uncertainty (\kms) associated with the \RV\ measurements in \gdr is\begin{equation}
\sigma_{V_{\mathrm R}}= \sqrt{(\sigma_{V_{\mathrm R}}^{\mathrm{MTA}})^{2}  + 0.11^2}.
\label{equ:rverror}
\end{equation}


In addition to \RV\, , other information is computed for each star by combining the single-transit information. This is not published in DR2, but is used to infer the quality of the \RV\ and to discard poor-quality measurements \citep{DR2-DPACP-54}. 
 
\section{Radial velocity results\label{sec:results}}
Various diagnostics are implemented in the automated verification {\tt AVMTA} (Fig.~\ref{fig:workflows}) to automatically verify the median radial velocity results produced by the {\tt MTA} for each star from the total of its transits. In the same way as for the single-transit radial velocities, a verification dataset is defined
that contains only stars for which we also have a ground-based radial velocity in the auxiliary external catalogues (Table \ref{tab:auxradvel}). The verification dataset covers the magnitude range of the RVS data (Fig.~\ref{fig:auxradvel-grvs}), but does not contain early-type or late-M stars. Table~\ref{tab:auxvalmta} lists the measurements performed by {\tt AVMTA}.


\begin{table}[h]
\begin{center}
\begin{adjustbox}{width=0.5\textwidth}
\begin{tabular}{l c  c  c  c }
  \hline\hline\\[-0.3cm]
  Cat Name & $Md$($\Delta$\RV) &$\sigma(\Delta$\RV) & $Md$($\sigma_{V_{\mathrm R}}^{\mathrm{MTA}})$ &$\sigma(\sigma_{V_{\mathrm R}}^{\mathrm{MTA}})$\\
   & \kms & \kms & \kms & \kms \\

  \hline\\[-0.3cm]
   CU6GB-cal &$-$0.01\,~~ &0.20 &0.14 & 0.07\\
     XHip &  0.35 & 0.51\ & 0.13 & 0.10 \\
  \hline\\[-0.3cm]
  CU6GB-val & $-$0.01\,~~& 0.51 & 0.23 & 0.23\\
  RAVE & 0.29& 1.03 & 0.37 & 0.27\\
  APOGEE &0.22 &0.89 & 0.71 & 0.67\\
  SIM & 0.27 &0.29& 0.23 & 0.13 \\
 \hline
\end{tabular}
 \end{adjustbox}
\caption{Median radial velocities compared to external catalogues. The units are \kms. The {\tt MTA} \RV\  (Eq. \ref{equ:rvmedian}) are compared with the external catalogues described in Sect. \ref{sssec:auxstd} and \ref{sssec:auxval}.
$Md$($\Delta$\RV) quantifies the overall systematic shifts between the RVS and the radial velocities of other catalogues.  $Md(\sigma_{V_{\mathrm R}}^{\mathrm{MTA}}$) quantifies the overall precision of the RVS radial velocities for the stars belonging to each catalogue. The precision is worse for the stars in APOGEE, which are the faintest.}
\label{tab:auxvalmta}
\end{center}
\end{table}

\subsection{Overall accuracy}

The accuracy indicates the systematic uncertainty. It is estimated using the verification dataset by calculating the difference between the RVS and the external catalogue measurements.
In Table \ref{tab:auxvalmta}, $Md$($\Delta V_{\rm R}$) is the median value of the residuals, calculated over the entire period covered by DR2 for the stars belonging to each catalogue in the verification dataset. The residuals $\Delta$\RV~=\RV$-$\RVref\ are the differences between the median radial velocity obtained by the pipeline (Eq. \ref{equ:rvmedian}) and those of the external catalogues (\RVref) and contain both the catalogue and the RVS uncertainties. They are in good agreement with the median residuals $Md$($\Delta$\RVtr) obtained with the single-transit radial velocities \RVtr and indicate the overall shift between the RVS and the other catalogues.

\begin{figure}[h]
\includegraphics[width=0.5\textwidth]{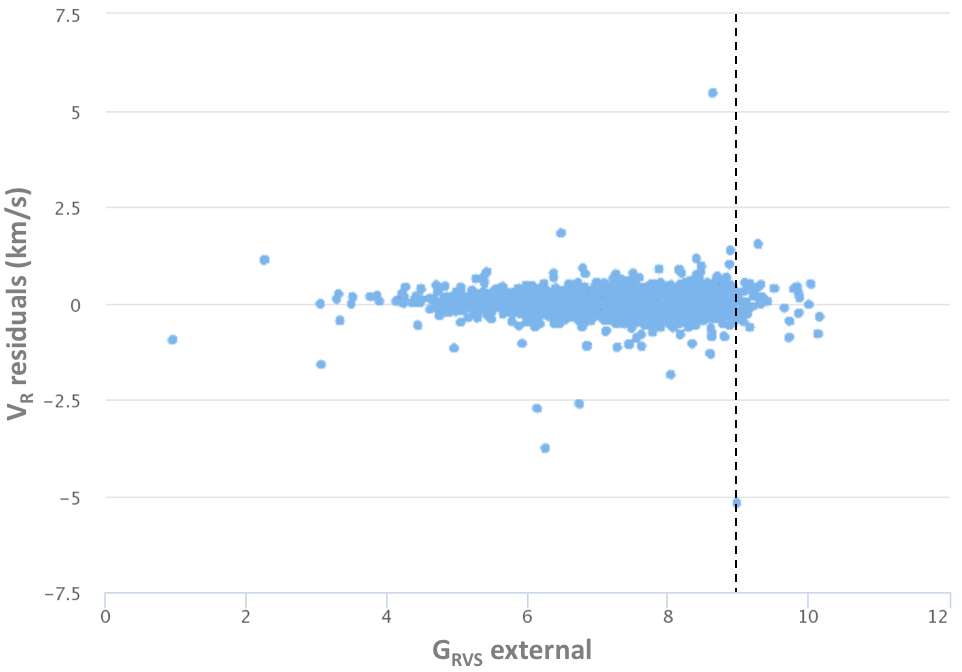}
\caption{Residuals $\Delta V_{\rm R}$ for the stars in CU6GB-cal as a function of \extgrvs. The overall median value is $-$0.01 \kms and the dispersion is 0.20 \kms (Table \ref{tab:auxvalmta}). The magnitude distribution of the CU6BG-cal stars is shown in Fig.~\ref{fig:auxradvel-grvs}. The stars with \extgrvs $\leq 9$ are used as standard stars in the wavelength calibration. The pipeline processes stars with \extgrvs $\leq 12$. There is no bright-end limit in the selection of the stars to process, but the pipeline removes the saturated spectra (Sect.~\ref{sssec:biascorr}).}
\label{fig:res-cu6GBcal}
\end{figure}

When we assume that the uncertainties in the external catalogue
are negligible, $Md(\Delta V_{\rm R}$) is an indicator of the overall accuracy (systematic zeropoint shift) in the RVS \RV. The low value for the stars in CU6GB indicates that overall, the RVS \RV\ zeropoint is in agreement with that of the standard stars, which implies that the pipeline processing did not introduce any significant systematic shift in general. Figure \ref{fig:res-cu6GBcal} shows the residuals as a function of \extgrvs\ for the CU6GB-cal stars, with no  evident magnitude-dependent zeropoint shift. 

The DR2 data do exhibit an overall shift of $\sim+300$~\ms compared
with the other catalogues (Table~\ref{tab:auxvalmta}). 
These shifts are acceptable for \gdr since the pre-launch end-of-mission requirement on systematic uncertainties was  $\leq 300$ \kms \citep[see][Table 1]{DR2-DPACP-46}.

\subsection{Overall precision}
The precision indicates random uncertainties. In Table~\ref{tab:auxvalmta} the overall precision of the RVS \RV\ measurements is quantified using the verification dataset with  $\sigma(\Delta V_{\rm R})$ and $ Md$($\sigma_{V_{\mathrm R}}^{\mathrm{MTA}}$) calculated over the 22 months covered by DR2. $\sigma(\Delta V_{\rm R})$ is the robust dispersion (see Eq.~\ref{eqn:robuststd}) of the residuals for the stars belonging to the different catalogues of the verification dataset. When we assume negligible uncertainties in the external catalogues, it is an indicator of the overall precision of the RVS measurements and is called external precision (this includes the precision of the external catalogue measurements). $Md$($\sigma_{V_{\mathrm R}}^{\mathrm{MTA}}$) is the overall internal precision, which is also shown in Table~\ref{tab:auxvalmta}, where $\sigma_{V_{\mathrm R}}^{\mathrm{MTA}}$ is estimated in the pipeline (Eq. \ref{equ:errormedian}). The internal precision is independent of the external catalogue uncertainty and depends on the dispersion of the RVS measurements. It also depends on the magnitude of the stars: it is higher for CU6GB-cal and XHip, which are dominated by bright stars (see Fig.~\ref{fig:auxradvel-grvs}) and lower for RAVE and APOGEE, which contain fainter stars. 

The CU6GB-cal \RVref\ are more precise ($\sigma < 0.1$; Table\ref{tab:auxradvel}) and are also expected to be more accurate than the other external catalogues. For these stars the agreement between the external and the internal precision is reached by adding 0.11 \kms \ in quadrature to the internal estimation. This shift is interpreted as a calibration floor. After the pipeline processing was completed, it was added quadratically to the MTA uncertainty (Eq. \ref{equ:rverror}) to improve their estimation
in a post-processing phase.

\begin{figure}[h]
\includegraphics[width=0.5\textwidth]{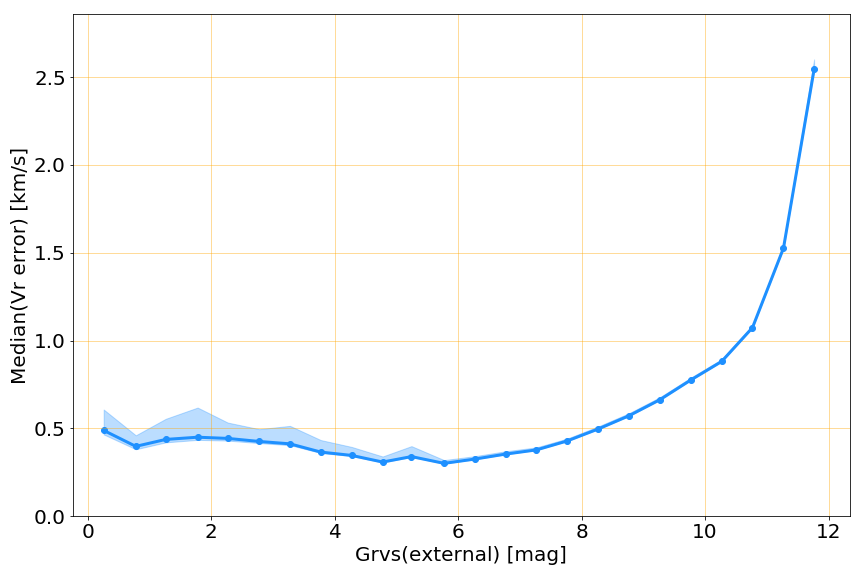}
\caption{Precision ($Md$($\sigma_{V_{\mathrm R}}^{\mathrm{MTA}}$) over magnitude bins) of the full \RV\ data set produced by the pipeline: 9\,816\,603 stars.  After an off-line validation campaign in which poor-quality \RV\ have been excluded ,$\sim$ 7.2 million stars are left for publication in DR2, and the internal precision was improved by $\sim$ 30\%}
\label{fig:precisionAll}
\end{figure}

Figure \ref{fig:precisionAll} shows the internal precision as a function of \extgrvs\ for the $\sim 10$ million \RV\  produced by the pipeline (not only those in the verification dataset: SB1 and variable stars are included). The median, $Md$($\sigma_{V_{\mathrm R}}^{\mathrm{MTA}}$), is calculated over 0.25 magnitude bins. The precision is even better than the pre-launch end-of-mission requirement of 1 \kms for stars brighter than \extgrvs $\sim 10.5$.

\subsection{Off-line validation}

When the processing of the data was completed and {\bf {\tt Automated Verification}} confirmed that the overall accuracy and precision of the \RV\  met the requirements, a validation campaign started on the $\sim 10$ million \RV\ produced by the pipeline. Its purpose was to analyse the global properties of the dataset and identify the \RV\ without sufficient quality to be published in DR2 \citep{DR2-DPACP-54}.
The exclusion criteria include those mentioned in previous sections, which are hot (\tempTeff\ $\geq 7000$ K) and cool stars (\tempTeff\ $\leq 3500$ K), see also Sec.~ \ref{sssec:detAP}, suspected SB2 detected as double-lined in 90\% of the transits, emission-line stars, and stars that are too faint (\intgrvs $ > 14$), stars for which the \RVtr\ has been detected as {\it ambiguous} (Sect. \ref{sssec:cu6spe_flagging}) in all transits. Excluded from DR2 are also the \RV\ for which \RVerr $\geq 20$ \kms, and $\sim$ 400 outliers \RV\ $\geq 500$ \kms that were obtained with poor-quality spectra.

After the above filters were applied, an exhaustive validation study of the \RV\ pipeline products was performed, which is described in the accompanying paper \citep{DR2-DPACP-54}. It includes an
analysis of the accuracy and the precision depending on the star nature, magnitude, number of transits, and sky distribution. Then, another validation campaign \citep{DR2-DPACP-39} was carried out on the entire DR2 data, including the products of the other pipelines, and resulted in the exclusion of some stars.
In the end, $\sim$ 2.6 million \RV\ were excluded, and the number of \gaia\ stars with a \RV\ in \gdr is $\sim$ 7.2 million.

\section{Conclusions} 

We have presented the spectroscopic pipeline that was used to process the \gaia\ RVS data to produce the radial velocity measurements
that are  released in the \gaia\ DR2.
The pipeline processed $\sim$ 280 million individual RVS spectra of stars with apparent \grvs\ magnitude brighter than $\sim 12$, distributed throughout the entire celestial sphere. 

We described the calibration of the RVS, the reduction of the spectra and the determination of the radial velocity \RV, together with the overall \RV\ performance estimated by the pipeline. The \RV\ recorded for each star is the median of all radial velocities obtained from individual observations of that star.
The radial velocity is measured through a fit of the RVS spectrum with a synthetic template spectrum. To select the appropriate template, a rough estimate of the stellar atmospheric parameters was performed by the pipeline. The large uncertainties affecting such estimates for the hottest (\Teff\ $\geq 7000$ K) and coolest stars (\Teff\ $\leq 3500$ K) implied large uncertainties in the resulting \RV\ because of template mismatch errors. The \RV\ of these stars were removed from DR2.

The overall  accuracy (systematic uncertainties) of the radial velocity products, estimated automatically through comparisons with ground-based radial velocity catalogues, is better than $ 0.3\,$\kms. 
The overall precision (random uncertainties), also estimated automatically using stars known to be stable from ground-based observations, is better than $1\,$\kms. The best precision of  $<$ 0.2\,\kms is obtained for the stars brighter than \grvs\ $\sim$ 7.5. 
Although this is an early stage of data processing of only 22 months of \gaia\ RVS data, the radial velocities we obtained already approach or exceed the pre-launch end-of-mission requirements for bright stars

\gdr contains the median \RV\ of $\sim$ 7.2 million stars. For bright stars, it provides the third component of the velocity vector, complementing the 2D proper motion information. 
The spectroscopic pipeline will be improved for the future data releases. This will include the \RV\ of fainter stars, more accurate and precise \RV\ for bright stars, variability information, rotational velocities, calibrated spectra, and at a later stage, individual-transit observation data.

\begin{acknowledgements}
We thank the referee, Johannes Andersen, for his constructive comments.

This work has made use of results from the European Space Agency (ESA) space mission {\it Gaia}, the
data from which were processed by the {\it Gaia} Data Processing and Analysis Consortium (DPAC).
Funding for the DPAC has been provided by national institutions, in particular the institutions
participating in the {\it Gaia} Multilateral Agreement. 
The {\it Gaia} mission website is
\url{http://www.cosmos.esa.int/gaia}. Most of the authors are current or past members of the ESA {\it Gaia}
mission team and of the {\it Gaia} DPAC and their work has been supported by 
the French Centre National de la Recherche Scientifique (CNRS), the Centre National
d' Etudes Spatiales (CNES), the L'Agence Nationale de la Recherche, the
R\'{e}gion Aquitaine, the Universit\'e de Bordeaux, the Utinam Institute of the Universit\'e
de Franche-Comt\'e, and the Institut des Sciences de l' Univers (INSU);
the Science and Technology Facilities Council and the United Kingdom Space
Agency; the Belgian Federal Science Policy Office (BELSPO) through various
Programme de D\'{e}veloppement d'Exp\'{e}riences Scientifiques (PRODEX) grants;
the German Aerospace Agency (Deutsches Zentrum fur Luft- und Raumfahrt
e.V., DLR); the Algerian Centre de Recherche 
en Astronomie, Astrophysique et G\'{e}ophysique of Bouzareah Observatory;
the Swiss State Secretariat for Education, Research,
and Innovation through the ESA PRODEX programme, the Mesures
d'Accompagnement, the Swiss Activit\'{e}s Nationales Compl\'{e}mentaires, and the
Swiss National Science Foundation; the Slovenian Research Agency (research
core funding No. P1-0188). 
This research has made use of the SIMBAD database \citep{2000A&AS..143....9W} developed and operated at CDS,
Strasbourg, France.

\end{acknowledgements}

\bibliographystyle{aa} 
\bibliography{refs} 

\clearpage
\onecolumn
\appendix

\section{Acronyms}\label{sect:acronyms}


\begin{longtable}{ll}
\caption{\label{tab:acronyms}Acronyms used in this paper.}\\
\hline\hline
Acronym & Description\\
\hline
\endfirsthead
\multicolumn{2}{c}%
{\tablename\ \thetable\ -- \textit{Continued from previous page.}}\\
\hline\hline
Acronym & Description\\
\hline
\endhead
\hline \multicolumn{2}{r}{\textit{Continued on next page.}}\\
\endfoot
\hline
\endlastfoot
AC   &   Across-scan (direction) \\
ADU&Analogue-to-digital unit \\
AGIS   &   Astrometric global iterative solution \\
AL   &   Along-scan (direction) \\
AP & Atmospheric parameters \\
AV&Automated verification \\
AVPP&Automated verification of pre-processing ({\it i.e.} ingestion workflow)\\
AVEXT&Automated verification of extraction ({\it i.e.} calibration preparation workflow)\\
AVFE&Automated verification of full extraction\\
AVSTA&Automated verification of STA\\
AVMTA &Automated verification of MTA\\
CCD   &   Charge-coupled device (detector) \\
CCF & Cross-correlation function \\
CPU&Central processing unit \\
CRNU&(CCD) Column response non-uniformity \\
CTI&Charge transfer inefficiency \\
CU6GB& Coordination Unit 6 ground-based radial velocity catalogue\\
CaU&Calibration unit \\
DR2   &   ({\it Gaia}) Data Release 2  \\
EPSL   &   Ecliptic pole scanning law \\
FL   &   First look \\
FoV&Field of view (also denoted FOV) \\
GACS&Gaia archive core systems \\
GSC23 & the second guide star catalogue version 2.3\\
IDT   &   Initial data treatment \\
IDU   &   Intermediate data update \\
IGSL   &   Initial {\it Gaia} source list \\
LSF&Line spread function \\
MTA&Multi-transit analysis \\
NIR & Near-infrared \\
NSL&Nominal scanning law \\
NU&Non-uniformity \\
OBMT&On-board mission timeline (in \gaia\ revolutions; 1 revolution = 6 hours)\\
OGA3&On-ground attitude from AGIS \\
PCA&Principal components analysis \\
PCF&Pearson correlation function\\
PEM   &   Proximity-electronics module (CCD) \\
PSF&Point spread function \\
RAVE&Radial velocity experiment \\
RMS&Root mean square \\
RP   &   Red photometer \\
RVDir & Radial velocity determination in direct space, using PCF\\
RVFou & Radial velocity determination with Fourier technique, using CCF\\
RVMDM & Radial velocity determination using minimum distance method \\
RVS   &   Radial Velocity Spectrometer \\
SAGA&System of accomodation of Gaia algorithms \\
SB2&Double-lined spectroscopic binary \\
SIM&Space interferometry mission \\
STA&Single-transit analysis \\
TDI&Time-delayed integration (CCD) \\
VO&Virtual object \\
VPU&Video processing unit \\
\end{longtable}

\end{document}